\documentclass[fleqn,usenatbib]{mnras}

\usepackage{newtxtext,newtxmath}

\usepackage[T1]{fontenc}
\usepackage{ae,aecompl}

\usepackage{graphicx}	
\usepackage{amsmath}	
\usepackage{tabularx}

\newcommand{\kms}{$\,$km$\,$s$^{-1}$}

\title[Formaldehyde Emission in NGC 7538 IRS1]{Thermal Formaldehyde Emission in NGC $\,$7538 IRS$\,$1}

\author[Shuvo et al.]{Onic I. Shuvo,$^{1,2}$\thanks{E-mail: oshuvo@masonlive.gmu.edu}
E. D. Araya,$^{2,3}$\thanks{E-mail: ed-araya@wiu.edu}
W. S. Tan,$^{2}$
P. Hofner,$^{3}$\thanks{Adjunct Astronomer at the National Radio Astronomy Observatory, 1003 Lopezville Road, Socorro, NM 87801, USA.}
\newauthor S. Kurtz,$^{4}$
Y. M. Pihlstr\"om,$^{5}$\thanks{Adjunct Astronomer at the National Radio Astronomy Observatory, 1003 Lopezville Road, Socorro, NM 87801, USA.} 
and I. M. Hoffman$^6$\\
\\
$^{1}$Department of Physics and Astronomy, George Mason University, 4400 University Drive, Fairfax, VA 22030, USA.\\
$^{2}$Physics Department, Western Illinois University, 
1 University Circle, Macomb, IL 61455, USA.\\
$^{3}$New Mexico Institute of Mining and Technology, 
Physics Department, 801 Leroy Place,
Socorro, NM 87801, USA.\\
$^{4}$Instituto de Radioastronom\'{\i}a y Astrof\'{\i}sica,
Universidad Nacional Aut\'onoma de M\'exico, Apdo. Postal 3-72, 
58090, Morelia, \\
Michoac\'an, Mexico.\\
$^{5}$Department of Physics and Astronomy, University of New Mexico, 210 Yale Blvd.\,NE, Albuquerque, NM 87131, USA.\\
$^{6}$Quest University, 3200 University Boulevard, Squamish, British Columbia V8B0N8, Canada.\\
}

\date{Accepted 2021 March 19. Received 2021 March 18; in original form 2020 October 16.}

\pubyear{2021}

\begin{document}
\label{firstpage}
\pagerange{\pageref{firstpage}--\pageref{lastpage}}
\maketitle
\begin{abstract}

Spectral lines from formaldehyde (H$_2$CO) molecules at cm wavelengths are typically detected in absorption and trace a broad range of environments, from diffuse gas to giant molecular clouds. In contrast, thermal emission of formaldehyde lines at cm wavelengths is rare. In previous observations with the 100$\,$m Robert C. Byrd Green Bank Telescope (GBT), we detected 2$\,$cm formaldehyde emission toward NGC$\,$7538 IRS1 -- a high-mass protostellar object in a prominent star-forming region of our Galaxy. We present further GBT observations of the 2$\,$cm and 1$\,$cm H$_2$CO lines to investigate the nature of the 2$\,$cm H$_2$CO emission. We conducted observations to constrain the angular size of the 2$\,$cm emission region based on a East-West and North-South cross-scan map. Gaussian fits of the spatial distribution in the East-West direction show a deconvolved size (at half maximum) of the 2$\,$ cm emission of 50\arcsec $\pm$ 8\arcsec. The 1$\,$cm H$_2$CO observations revealed emission superimposed on a weak absorption feature. A non-LTE radiative transfer analysis shows that the H$_2$CO emission is consistent with quasi-thermal radiation from dense gas ($\sim 10^5\,$to $10^6\,$cm$^{-3}$). We also report detection of 4 transitions of CH$_3$OH (12.2, 26.8, 28.3, 28.9$\,$GHz), the (8,8) transition of NH$_3$ (26.5$\,$GHz), and a cross-scan map of the 13$\,$GHz SO line that shows extended emission ($> 50$\arcsec).
\end{abstract}




\section{Introduction}

Formaldehyde is a tracer of high density gas in high-mass star forming regions and it is a reliable density probe in Galactic molecular clouds (e.g., \citealt{Mangum_2008ApJ...673..832M}, \citealt{Ginsburg_2011ApJ...736..149G}). The rotational levels of ortho-formaldehyde (H$_2$CO) are split in doublets, commonly known as K-doublets. The K-doublet lines from the three lowest H$_2$CO rotational energy levels correspond to wavelengths of 6, 2, and 1$\,$cm (e.g., \citealt{Mangum_1993ApJS...89..123M}, \citealt {Thaddeus_1972ApJ...173..317T}).

While thermal emission of high-frequency formaldehyde transitions has been detected in high-mass star forming regions (e.g., \citealt{Ceccarelli_2003A&A...410..587C}), thermal emission of the lowest K-doublet transitions is rare (e.g., \citealt{Evans_1975ApJ...199..383E}; \citealt{Araya_2006AJ....132.1851A}). At present, thermal 6$\,$cm H$_2$CO emission has been detected only toward the Orion BN/KL region (e.g., \citealt{Araya_2006ApJ...643L..33A}). Galactic quasi-thermal emission of the 2$\,$cm transition has been reported toward a few sources (Orion-KL, OMC-2, $\rho$ Oph B, DR 21(OH), and three regions in W51; e.g, \citealt{Ginsburg_2016A&A...595A..27G}, \citealt{Johnston_1984ApJ...285L..85J}, and references therein).\footnote{We use the term {\it quasi-thermal} to refer to gas in between a state dominated by non-LTE excitation resulting in anomalous absorption  ($T_{ex} \, < \, T_{CMB}$ in the low-density regime; e.g., \citealt{Evans_1975ApJ...196..433E_Anomalous}) and LTE excitation dominated by molecular collisions (high-density regime).} Thermal emission at 1$\,$cm has been found toward 17 sources \citep{McCauley_2011ApJ...742...58M}. In contrast, formaldehyde masers have only been unambiguously detected in the 6$\,$cm transition (e.g., \citealt{Chen_2017ApJ...851L...3C}, \citealt{Ginsburg_2015A&A...584L...7G}, \citealt{Araya_2007IAUS..242..110A}, \citealt{Hoffman_2003ApJ...598.1061H})\footnote{\cite{Chen_2017MNRAS.466.4364C} reported possible 2$\,$cm H$_2$CO emission lines toward G23.01$-$0.41 and G29.96$-$0.02, in addition to NGC$\,$7538; see below.}. Currently, 10 Galactic regions are know to harbor 6$\,$cm masers \citep{Andreev_2017ApJS..232...29A, Chen_2017ApJ...851L...3C}, one of them is NGC$\,$7538 -- the region where 6$\,$cm H$_2$CO masers were first confirmed (see review by \citealt{Araya_2007IAUS..242..110A}).

NGC$\,$7538 is an active site of high-mass star formation located at a distance of $\,$2.65$\pm$0.12$\,$kpc \citep{Moscadelli_2009ApJ...693..406M}. This region is one of the richest maser sources known (e.g., \citealt{Galvan-Madrid_2010ApJ...713..423G}). Over a dozen different molecular maser transitions have been found in NGC$\,$7538, including one of the rare 6$\,$cm H$_{2}$CO masers (e.g., \citealt{Araya_2007ApJS..170..152A}). 

Using the 100$\,$m Robert C. Byrd Green Bank Telescope (GBT), we detected 2$\,$cm H$_{2}$CO emission toward NGC$\,$7538 IRS$\,$1 \citep{Andreev_2017ApJS..232...29A}. This emission was independently detected by \cite{Chen_2017MNRAS.466.4364C}. Given the non-detection of a 2$\,$cm maser in interferometric observations \citep{Hoffman_2003ApJ...598.1061H}, \cite{Chen_2017MNRAS.466.4364C} argued that the 2$\,$cm H$_2$CO emission in NGC$\,$7538 IRS$\,$1 is a variable maser, noting that variability has been detected in the 6$\,$cm H$_2$CO maser in the same source (e.g., \citealt{Andreev_2017ApJS..232...29A}). 

In this work, we present GBT observations of the 2$\,$cm and 1$\,$cm H$_2$CO transitions to investigate the nature of the 2$\,$cm emission\footnote{Preliminary results from this work were presented at a conference by \cite{Yuan_2011AAS...21812904Y}.}. Depending on whether the emission is thermal or maser, the line would provide information on completely different scales, from tens of a.u.'s in the case of masers, to sub-pc structures in the case of thermal emission. In particular, if the 2$\,$cm H$_2$CO emission is thermal, it would trace a connection between the large-scale molecular cloud in NGC$\,$7538 and the molecular core hosting the IRS1 star formation site.

\section{Observations}

\subsection{2 cm H$_2$CO}

Observations of the 2$\,$cm H$_2$CO transition in NGC$\,$7538 were conducted with the GBT in November 2008. We observed the J$_{KaKc}$=2$_{11}$-2$_{12}$ transition of formaldehyde ($\nu_0$=14.488480$\,$GHz, H$_2$CO 2$\,$cm line)\footnote{\label{fn1}Spectroscopy information in this paper is from Splatalogue (http://www.cv.nrao.edu/php/splat/) and the NIST Lovas catalog (https//physics.nist.gov/cgi-bin/micro/table5/start.pl), unless indicated otherwise.} with a bandwidth (BW) of 12.5$\,$MHz observed in frequency switching mode (5 minutes per scan) that resulted in an effective bandwidth of 6.25$\,$MHz ($\sim$130$\,$km s$^{-1}$), 9 level sampling, 8192 channels and final channel separation of 21.3$\,$kHz (0.442$\,$km s$^{-1}$) after smoothing. Given the capabilities of the GBT spectrometer, we conducted observations using additional independent bands (spectral windows) tuned to the $J_K = 2_{0}$-3$_{-1}$ \emph{E} transition of CH$_3$OH (12.178593$\,$GHz) and the $N_J = 2_{1}$-$1_{1}$ transition of SO (13.043814$\,$GHz). The system temperature was approximately 30$\,$K. A calibration diode was used to set the antenna temperature scale, and flux density calibration was done using the telescope gain values provided by the observatory in the {\sc GBTIDL}\footnote{{\sc IDL} (Interactive Data Language) is a trademark of Harris Geospatial Corp.} software package. We observed 3C48 in position switching mode to check the flux density calibration; we measured $S_\nu = 1.9\,$Jy at 14.8$\,$GHz, which agrees within 10\% with the expected value of 1.74$\,$Jy\footnote{\label{vlaflux} https://www.vla.nrao.edu/astro/calib/manual/fluxscale.html}. Based on the pointing observations of 3C48, we measured a half-power beam width (HPBW) of 52\arcsec~at the frequency of the 2$\,$cm H$_2$CO transition.

To investigate the angular size of the 2$\,$cm H$_2$CO emission region and measure its brightness temperature, we obtained a 13-point (East-West, North-South) cross-scan map; the pointings were offset in the RA and Dec directions by $\pm 27$\arcsec, $\pm 54$\arcsec, and $\pm 108$\arcsec~with respect to the location of NGC$\,$7538 IRS1 (R.A= 23$^{h}$13$^{m}$45.36$^{s}$ and Dec= +61$^{\circ}$ 28\arcmin 10.45\arcsec, J2000)\footnote{Given the HPBW of the telescope, our IRS1 pointing also includes IRS2 and IRS3, e.g., \cite{Akabane_1992PASJ...44..421A}.}.

\subsection{1$\,$cm H$_2$CO}\label{sec:obs_1cm}

We also observed the J$_{KaKc}$=3$_{12}$-3$_{13}$ transition of formaldehyde ($\nu_0$=28.974804$\,$GHz, 1$\,$cm H$_2$CO line) with the GBT on 30 October 2008. The spectrometer was used with a 50$\,$MHz BW ($\sim$500$\,$km s$^{-1}$), 9 level sampling, 8192 channels and final channel separation of 18.3$\,$kHz (0.190$\,$km$\,$s$^{-1}$) after smoothing. In the spectral window of the 1$\,$cm H$_2$CO line, the $J_K = 8_{2}$-9$_{1}$ \emph{A$^-$} transition of CH$_3$OH (28.969942$\,$GHz) was also included\footnote{Rest frequency from the JPL catalog (https://spec.jpl.nasa. gov/ftp/pub/catalog/catform.html) as listed at the time of the observations, which was also used by \cite{McCauley_2011ApJ...742...58M}. This frequency differs by 12$\,$kHz with respect to the rest frequency reported in the Lovas catalog.}. Three additional spectral windows were used to simultaneously observe the (J,K) = (8,8) NH$_3$ (26.518981$\,$GHz) line, and two transitions 12$_{2}$-12$_{1}$ \emph{E} (26.847205$\,$GHz) and 4$_{0}$-3$_{1}$  \emph{E} (28.316031$\,$GHz) of CH$_3$OH. The HPBW of the telescope at 29$\,$GHz is $\sim$26\arcsec.

The observation procedure was position-switching (ON-OFF mode) with 2 minutes on both ON and OFF (reference) positions per scan. We obtained 12 scans of NGC$\,$7538$\,$IRS1. The reference position was set to match the azimuth-elevation path tracked by the telescope during the corresponding 2 minutes ON-source observations. The system temperature was approximately 45$\,$K. Calibration and data reduction were done using GBTIDL. We observed the quasars 3C48 and 2148+6107 for pointing and system checking. We measured a 3C48 flux density of $S_\nu = 0.92\,$Jy at 29$\,$GHz, which agrees within 12\% with the expected value of 0.82$\,$Jy\textsuperscript{\ref{vlaflux}}. The pointing errors are estimated to be $\sim 5$\arcsec~in both RA and Dec based on the observations of 2148+6107.

\section{Results}

\begin{figure}
\includegraphics[trim=4.5cm 9.8cm 8cm 5cm, clip, width = \columnwidth]{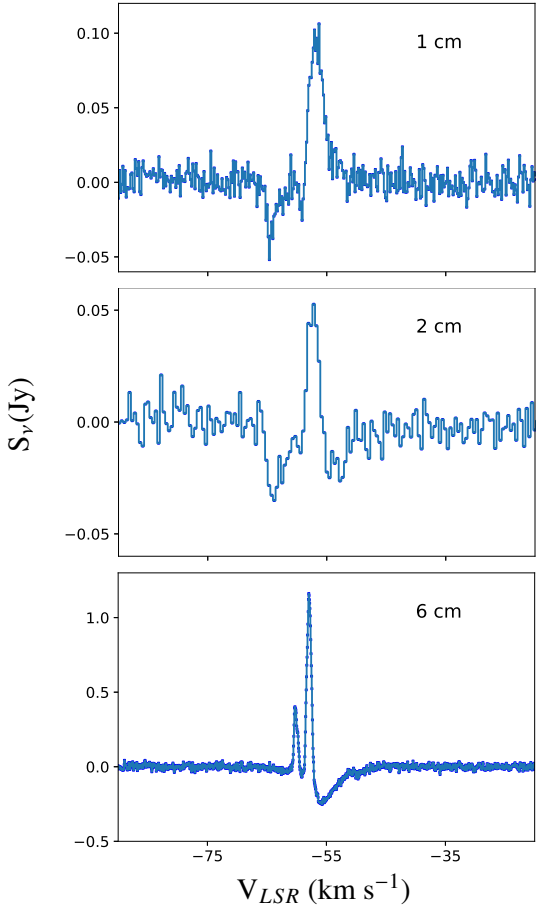}
\caption{Spectra of 1$\,$cm, 2$\,$cm and 6$\,$cm H$_2$CO transitions toward NGC$\,$7538$\,$IRS1. Note the detection of emission and absorption in all transitions. The data shown in the figure are available as Supporting Information.}
\label{fig:figure_1cm_2cm_6cm}
\end{figure}

\begin{figure}
\includegraphics[trim=4.5cm 5.6cm 8cm 5cm, clip, width = \columnwidth]{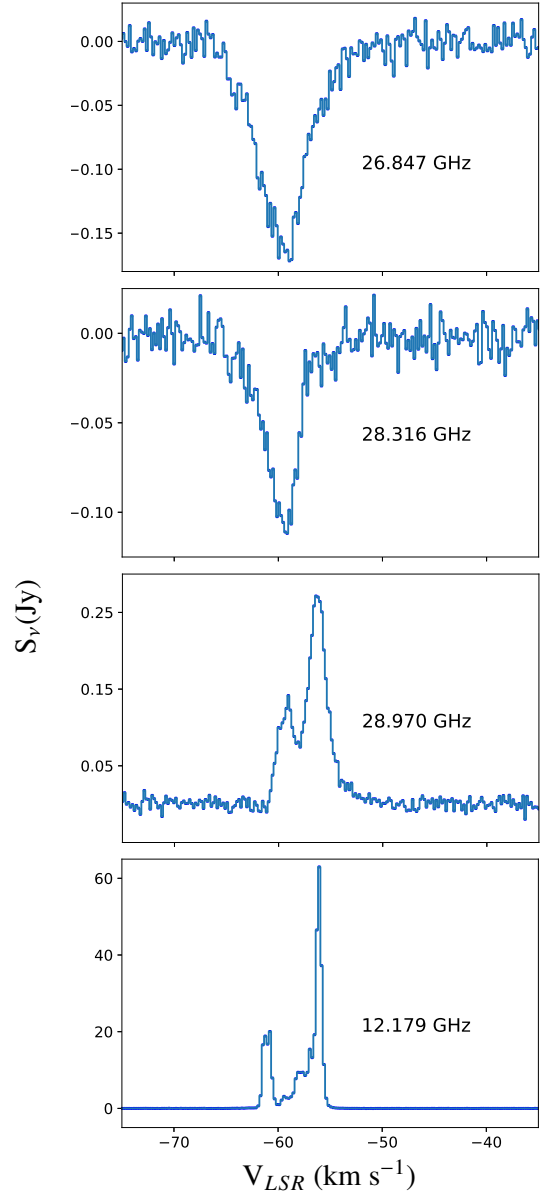}
\caption{Detection of CH$_3$OH transitions toward NGC$\,$7538$\,$IRS1. Two transitions show absorption (top panels; 26.847$\,$GHz and 28.316$\,$GHz) and two transitions show emission lines (lower panels; 28.970$\,$GHz and 12.178$\,$GHz). The data shown in the figure are available as Supporting Information.}
\label{fig:figure_CH3OH_26GHz_28GHz}
\end{figure}

\begin{figure}
\includegraphics[trim=4.5cm 9.2cm 8cm 8.6cm, clip, width = \columnwidth]{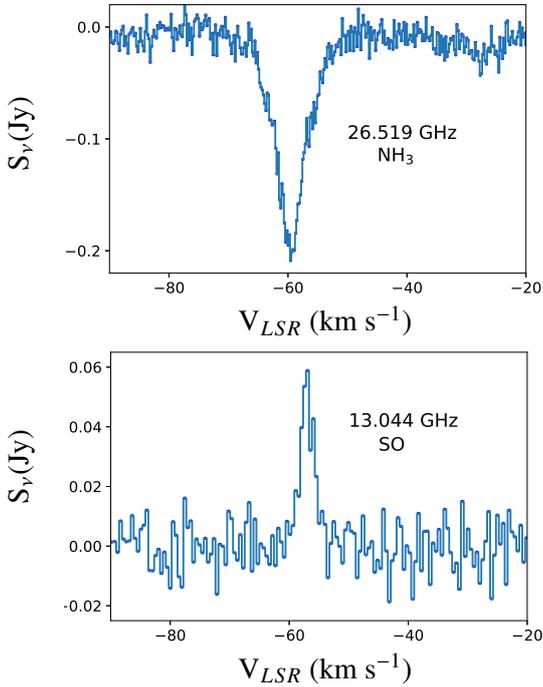}
\caption{Detection of NH$_3$ absorption and SO emission in NGC$\,$7538$\,$IRS1. The data shown in the figure are available as Supporting Information.}
\label{fig:figure_NH3_SO}
\end{figure}

\begin{figure*}
	\includegraphics[trim=1.8cm 5.2cm 1cm 7cm, clip, width=\textwidth]{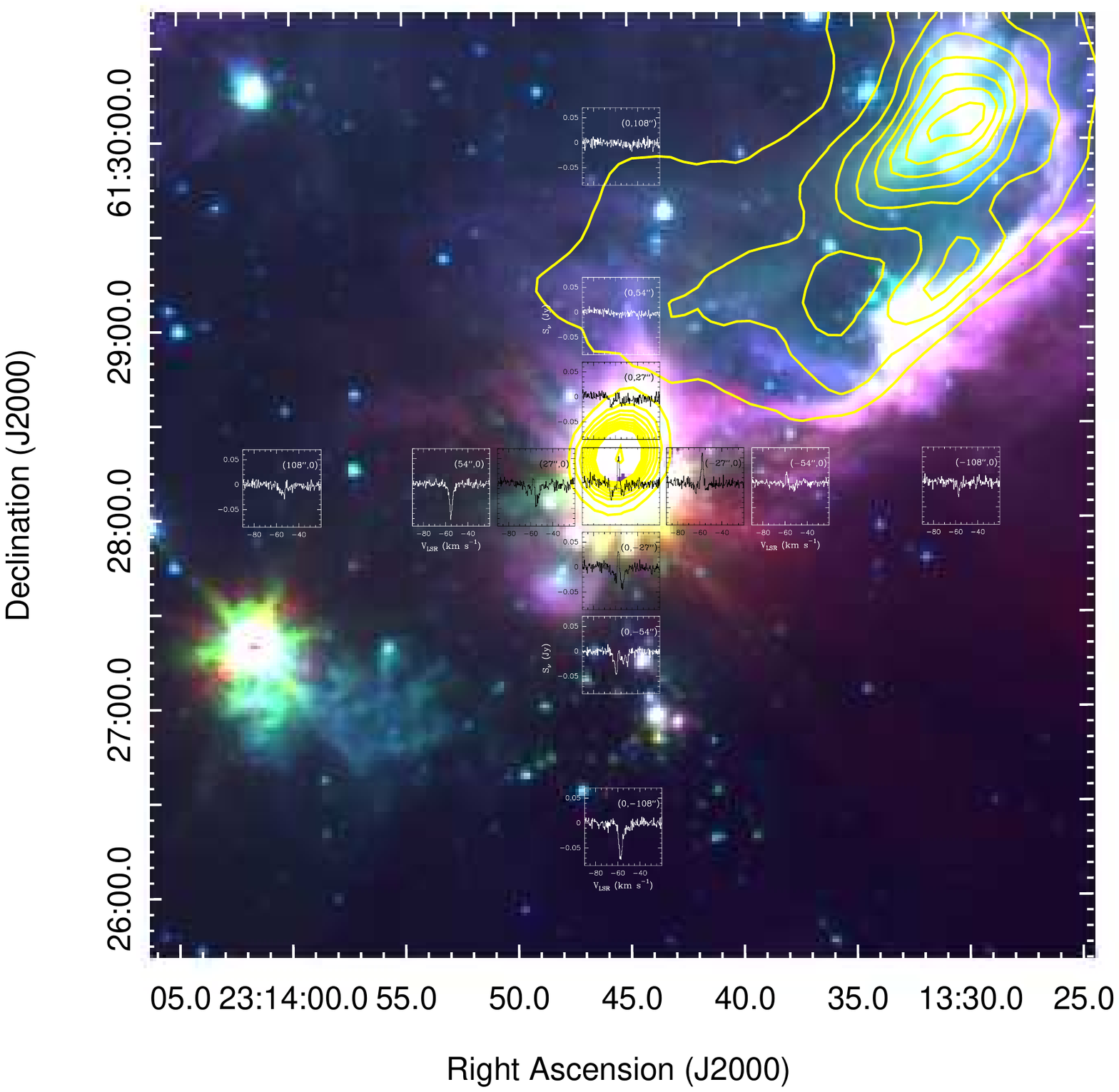}
	\centering
    \caption{Cross-scan map of 2$\,$cm H$_2$CO obtained with the GBT. The spectra are superimposed with a three-color (\textcolor{blue}{3.6 $\mu$m}, \textcolor{green}{4.5 $\mu$m} and \textcolor{red}{5.8 $\mu$m}) Spitzer/IRAC image of NGC$\,$7538. The central position of the cross scan map is R.A=23$^{h}$13$^{m}$45.36$^{s}$and Dec=+61$^{\circ}$28\arcmin 10.45\arcsec (J2000). Each spectrum box is 25\arcsec $\times$ 25\arcsec~in size (high-resolution images of the spectra are shown in Figure~\ref{fig:figure_H2CO_14GHz}). The HPBW of the GBT at 2$\,$cm is 52\arcsec, thus, the offsets from the central position are half-beam, full-beam, and two-beam spacing in the E-W and N-S directions. The yellow contours show 6$\,$cm VLA continuum from the NRAO image archive. The contour levels start at 39 $\,$mJy$\,$beam$^{-1}$ and continuing to 1104 $\,$mJy$\,$beam$^{-1}$ (peak intensity) in steps of 59 $\,$mJy$\,$beam$^{-1}$ $\times$ (1, 2, 3, 4, 5, 6, 7, 8, 17, 18). Note that four pointing positions [($-$108\arcsec,0),(0,$-$108\arcsec),(54\arcsec,0),(108\arcsec,0)] do not show strong evidence of overlapping emission and absorption lines.}
    \label{fig:map_figure}
    
\end{figure*}

Figure~\ref{fig:figure_1cm_2cm_6cm} shows the emission and absorption spectra of the three lowest K-doublet transitions of H$_2$CO towards NGC$\,$7538$\,$IRS1. The 1$\,$cm and 2$\,$cm H$_2$CO spectra are from the observations reported in this work, while the 6$\,$cm H$_2$CO spectrum is from GBT observations reported by \cite{Araya_2007ApJS..170..152A}. Spectra of the other molecular transitions observed in this work are shown in Figures~\ref{fig:figure_CH3OH_26GHz_28GHz} and \ref{fig:figure_NH3_SO}.

As mentioned above, we conducted cross-scan observations of 2$\,$cm H$_2$CO, 12.2$\,$GHz CH$_3$OH and 13.044$\,$GHz SO transitions, simultaneously. The spectra of the 2$\,$cm H$_2$CO cross-scan are shown in Figure~\ref{fig:map_figure}. To explore the distribution of 2$\,$cm H$_2$CO in the context of the IR and radio continuum environment in NGC$\,$7538$\,$IRS1, we show the 2$\,$cm H$_2$CO spectra superimposed with a Spitzer/IRAC\footnote{https://irsa.ipac.caltech.edu/data/SPITZER/Enhanced/SEIP/} image and 6$\,$cm VLA continuum from the NRAO image archive\footnote{http://www.aoc.nrao.edu/$\sim$vlbacald/ArchIndex.shtml}.

\begin{figure*}
\includegraphics[trim=3cm 2cm 2cm 1.8cm, clip,scale = 0.9]{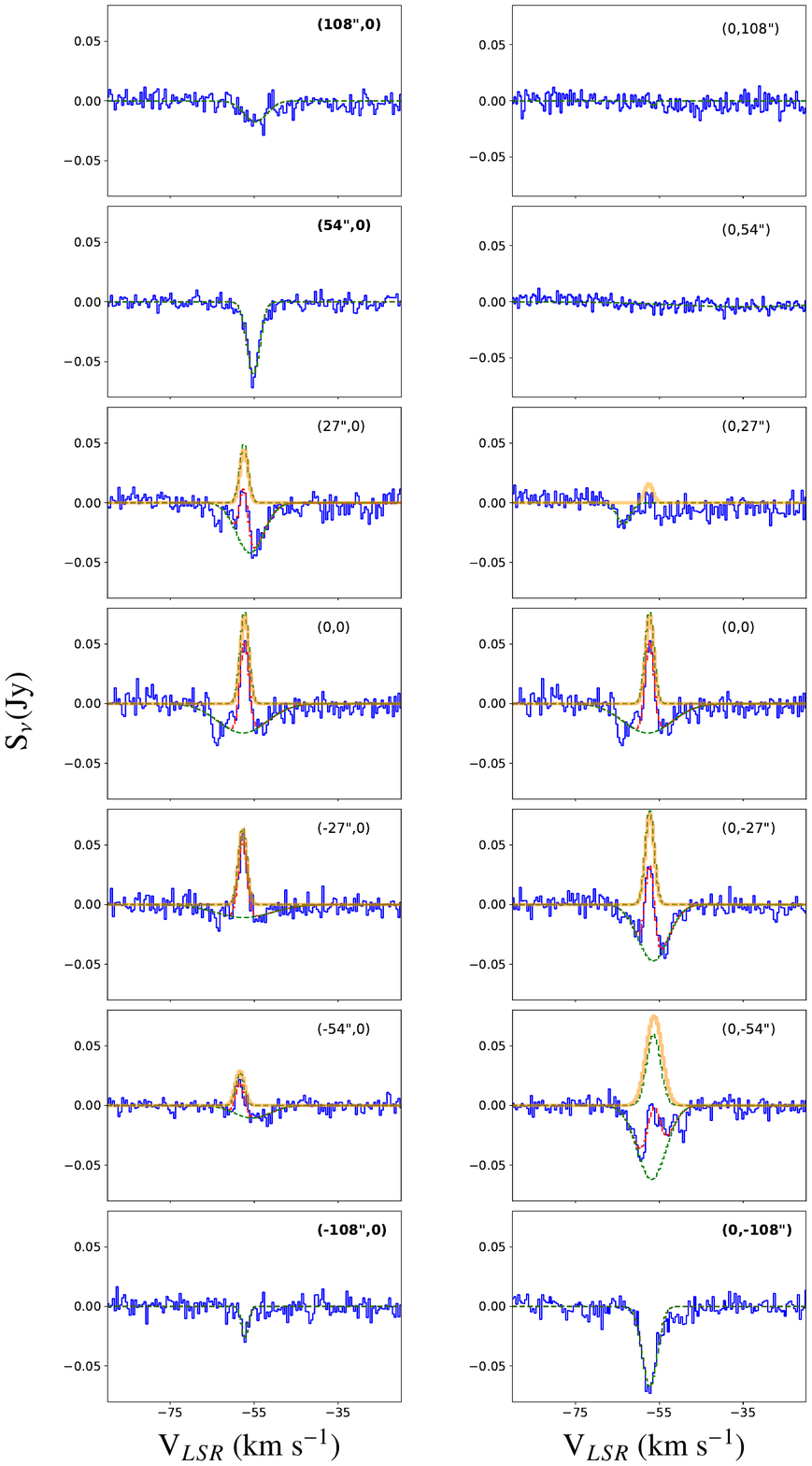}
\caption{Spectra of 2$\,$cm H$_2$CO emission and absorption toward NGC$\,$7538. Gaussian fits are included as dashed lines (green curves show individual components, red curves show the combined fit when more than one component was fit). The East-West spectra are shown in the left column, the North-South pointings are shown in the right column. Note that the same central pointing spectrum is shown in both the left and right columns. The highlighted spectra (bold coordinates) were used to interpolate the H$_2$CO absorption velocities (see Section~\ref{sec:2cm_abs} and Figure~\ref{fig:map_figure}). Thick solid (orange) lines are Gaussian fits of the emission lines after constraining the peak absorption velocities (see Section~\ref{sec:2cm_abs}). The data shown in the figure are available as Supporting Information.}
\label{fig:figure_H2CO_14GHz}
\end{figure*}

\begin{figure*}
\includegraphics[trim=3cm 2cm 2cm 1.8cm, clip,scale = 0.9]{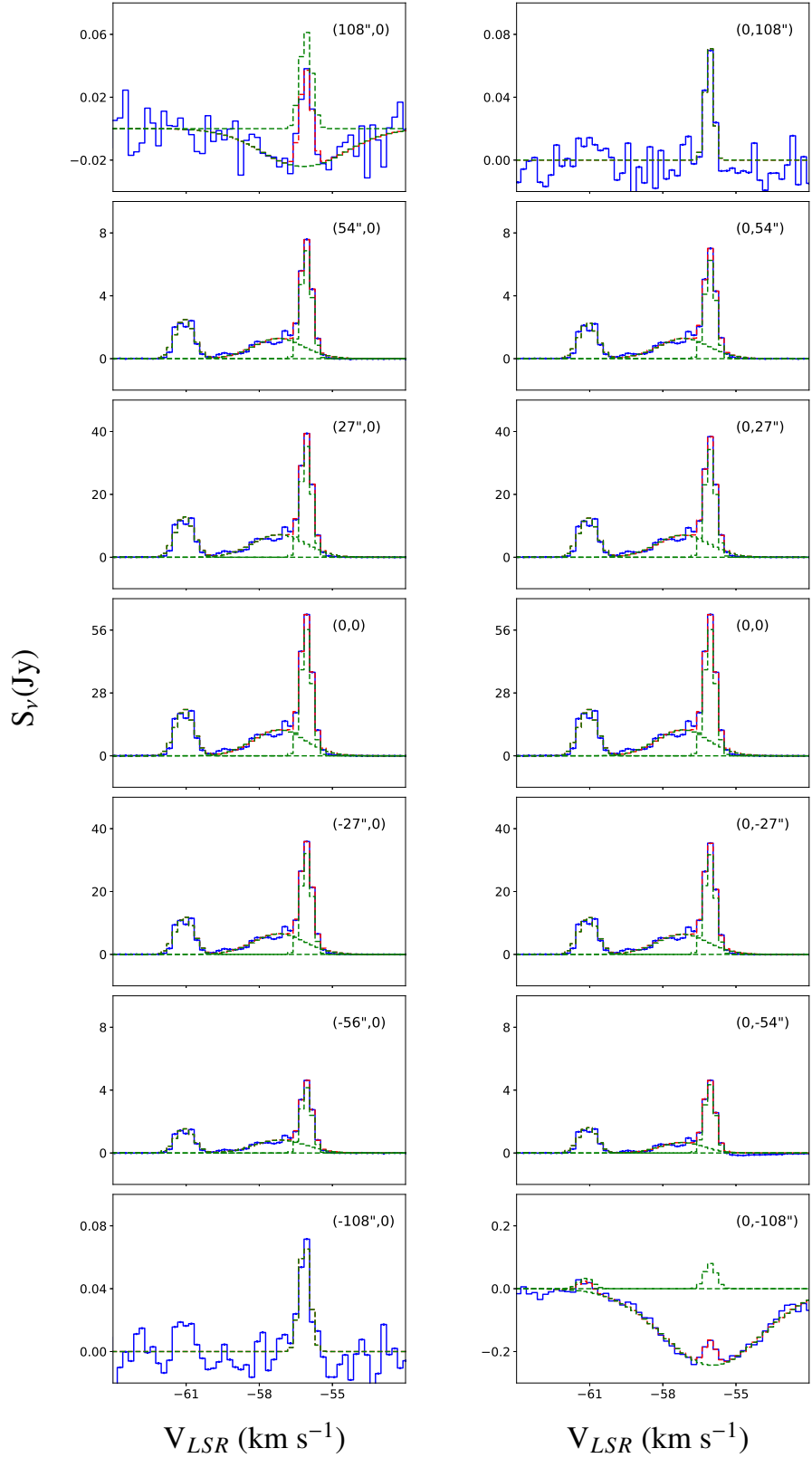}
\caption{Same as Figure~\ref{fig:figure_H2CO_14GHz} but for 12.2$\,$GHz CH$_3$OH. We show the same velocity range in all panels for consistency; however, the velocity range was optimized for emission and not for the (108\arcsec,0) and (0,$-$108\arcsec) pointings that show significant absorption. The two absorption spectra are shown with a larger velocity range in Figure~\ref{fig:figure_CH3OH_12GHZ_Abs}. The data shown in the figure are available as Supporting Information.}
\label{fig:figure_CH3OH_12GHz_EW_NS}
\end{figure*}

\begin{figure}
\includegraphics[trim=4.5cm 11cm 8cm 6cm, clip, width = \columnwidth]{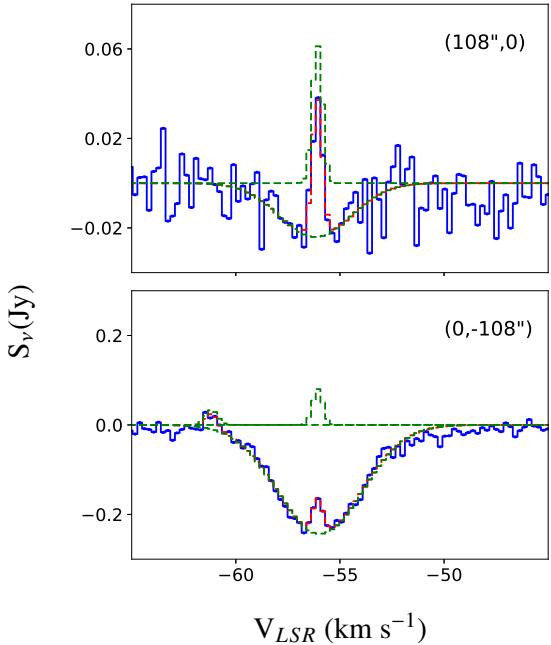}
\caption{Spectra and Gaussian fits for the two pointing positions with prominent 12.2$\,$GHz CH$_3$OH absorption as shown in Figure~\ref{fig:figure_CH3OH_12GHz_EW_NS}, i.e., (108\arcsec,0) and (0, $-$108\arcsec). The data shown in the figure are available as Supporting Information.}
\label{fig:figure_CH3OH_12GHZ_Abs}
\end{figure}

\begin{figure*}
\includegraphics[trim=3cm 2cm 2cm 1.8cm, clip,scale = 0.9]{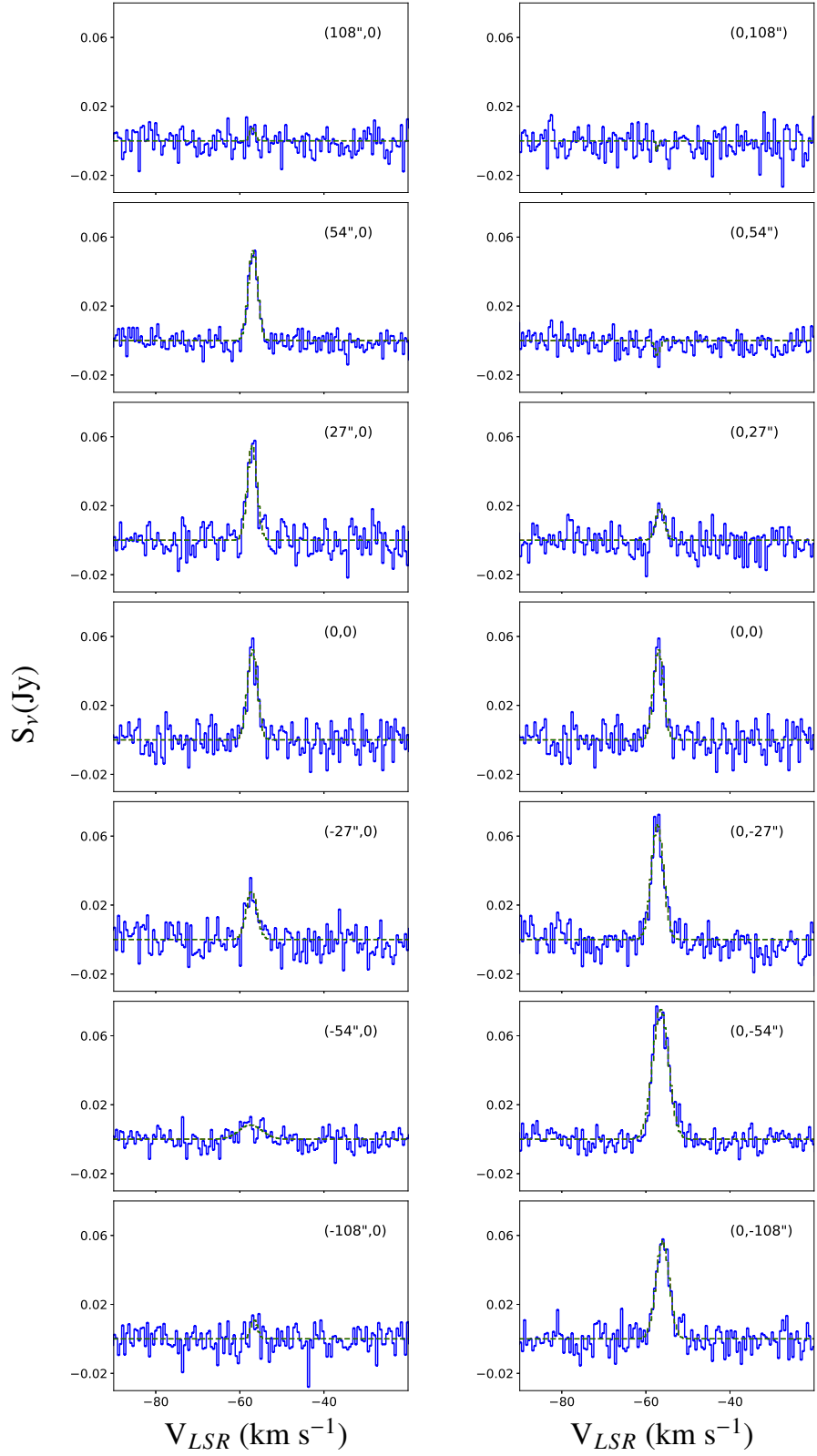}
\caption{Same as Figure~\ref{fig:figure_H2CO_14GHz} but for 13$\,$GHz SO. The data shown in the figure are available as Supporting Information.}
\label{fig:figure_SO_13GHZ}
\end{figure*}

The detections were fit with Gaussian profiles; Figures~\ref{fig:figure_H2CO_14GHz} to \ref{fig:figure_SO_13GHZ} show the Gaussian fits of all 13 cross-scan pointings (East-West, North-South) of the 2$\,$cm H$_2$CO,  12.2$\,$GHz CH$_3$OH, and 13.044$\,$GHz SO observations. The line parameters of 2$\,$cm H$_2$CO observations from the free fit are listed in Table~\ref{tab:H2CO}, whereas the line parameters from the fit after constraining the peak absorption velocities are listed in Table~\ref{tab:H2CO_EW_NS_after_constrain} (see Section~\ref{sec:2cm_abs} for details). The free fit line parameters of  12.2$\,$GHz CH$_3$OH, and 13.044$\,$GHz SO observations are listed in Tables~\ref{tab:CH30H}, and \ref{tab:SO} respectively. In the case of 12.2$\,$GHz CH$_3$OH, absorption was detected only in the pointing positions (108\arcsec,0) and (0, $-$108\arcsec); we highlight the absorption spectra in Figure~\ref{fig:figure_CH3OH_12GHZ_Abs}. Table~\ref{tab:H2CO_CH3OH_NH3} lists the line parameters of other transitions of H$_2$CO, CH$_3$OH, and NH$_3$.

We note that \cite{McCauley_2011ApJ...742...58M} also report GBT observations of the 1$\,$cm H$_2$CO line toward NGC$\,$7538$\,$IRS1. At a first glance, the spectrum in their Figure~1 shows a very similar line profile to the one shown in our Figure~\ref{fig:figure_1cm_2cm_6cm}. However, converting their peak $T_A^*$ to flux density (following the flux calibration notes in their Section~3) results in a value lower than ours by a factor
of $\sim$0.4 (the same is obtained if we calibrate our data directly to $T_A^*$ in GBTIDL). We argue that the line parameter values reported in this work (Table~\ref{tab:H2CO_CH3OH_NH3}) are more reliable than the values reported by \cite{McCauley_2011ApJ...742...58M} in their Table~2 because: 1) as explained in Section~\ref{sec:obs_1cm}, we observed the NRAO K. Jansky Very Large Array (VLA) flux density calibrator 3C48 to independently check our flux density scale and our measurement agrees within 12\% with the expected value\footnote{Note that our observations were conducted years before the 3C48 flare began in 2018; https://science.nrao.edu/facilities/vla/ docs/manuals/oss/performance/fdscale.}, 2) the line parameters of the 1$\,$cm H$_2$CO line reported in Table~2 of \cite{McCauley_2011ApJ...742...58M} show that the line was fit with two overlapping Gaussians, one absorption and one emission line, but both with approximately the same absolute intensity (0.1$\,$K), which disagrees with the line profile of the 1$\,$cm H$_2$CO line where the emission clearly has a greater line intensity than the absorption (e.g., Figure~\ref{fig:figure_1cm_2cm_6cm}). By assuming that both emission and absorption features have similar absolute intensities, the linewidths of the overlapping Gaussians reported by \cite{McCauley_2011ApJ...742...58M} were significantly broader than what the spectrum suggests.

\begin{table*}
    \centering
	\caption{Line parameters of 2$\,$cm H$_2$CO observations.}
	\label{tab:H2CO}
	\begin{tabular}{lrrrrr} 
		\hline
		Position    & S$_\nu$~~~~~~~ & RMS          & V$_{LSR}$~~~~~      & Width~~         &$\int$S$_\nu$dv~~~~~~\\
		 ($\Delta$RA, $\Delta$Dec)          & (Jy)~~~~~~  & (Jy)~        &(km s$^{-1}$)~~~ & (km s$^{-1}$) & (Jy km s$^{-1}$)\\
		\hline
NGC7538 IRS1        & 0.077(0.005)     & 0.006 & $-$57.34(0.07)   & 2.5(0.2)   & 0.21(0.03)\\
                    & $-$0.025(0.003)  & 0.006 & $-$57.75(0.52)   & 13.6(1.5)  & $-$0.36(0.09)\\
(108\arcsec,0)      & $-$0.018(0.002)  & 0.007 & $-$54.69(0.38)   & 6.5(0.9)   & $-$0.12(0.03)\\
(54\arcsec,0)       & $-$0.061(0.002)  & 0.004 & $-$55.19(0.07)   & 3.6(0.2)   & $-$0.23(0.02)\\
(27\arcsec,0)       & 0.049(0.006)     & 0.008 & $-$57.47(0.12)   & 2.4(0.4)   & 0.13(0.04)\\
                    & $-$0.042(0.006)  & 0.008 & $-$56.01(0.32)   & 7.7(0.6)   & $-$0.34(0.08)\\
($-$27\arcsec,0)    & 0.064(0.004)     & 0.006 & $-$57.76(0.07)   & 2.4(0.2)   & 0.17(0.02)\\
                    & $-$0.011(0.002)  & 0.006 & $-$57.50(1.10)     & 15.8(3.0)  & $-$0.18(0.07)\\
($-$54\arcsec,0)    & 0.027(0.003)     & 0.004 & $-$58.39(0.12)   & 2.4(0.3)   & 0.07(0.02)\\
                    & $-$0.011(0.002)  & 0.004 & $-$55.46(0.86)   & 10.0(1.5)  & $-$0.11(0.04)\\
($-$108\arcsec,0)   & $-$0.026(0.004)  & 0.006 & $-$57.22(0.15)   & 1.9(0.4)   & $-$0.05(0.02)\\
(0,108\arcsec)      & \ldots           & 0.006 & \ldots           & \ldots     & \ldots \\
(0,54\arcsec)       & \ldots           & 0.005 & \ldots           & \ldots     & \ldots \\
(0,27\arcsec)       & $-$0.018(0.004)  & 0.007 & $-$63.51(0.38)   & 3.7(0.9)   & $-$0.07(0.03)\\
(0,$-$27\arcsec)    & 0.079(0.006)     & 0.006 & $-$57.29(0.06)   & 2.6(0.2)   & 0.21(0.03)\\
                    & $-$0.047(0.006)  & 0.006 & $-$56.48(0.21)   & 8.4(0.6)   & $-$0.42(0.08)\\
(0,$-$54\arcsec)    & 0.060(0.019)     & 0.005 & $-$56.37(0.14)   & 3.9(0.7)   & 0.25(0.12)\\
                    & $-$0.062(0.019)  & 0.005 & $-$56.85(0.22)   & 8.0(0.9)   & $-$0.53(0.22)\\
(0,$-$108\arcsec)   & $-$0.067(0.003)  & 0.006 & $-$57.27(0.09)   & 4.2(0.2)   & $-$0.30(0.03)\\
		\hline
		\multicolumn{6}{p{10cm}}{Line parameters obtained from Gaussian fits, 1$\sigma$ statistical errors from the fit are reported in parenthesis. The spectra were smoothed to a channel width of 0.442$\,$km s$^{-1}$.}\\	
	\end{tabular}
\end{table*}

\begin{table*}
	\centering
	\caption{Line parameters of 2$\,$cm H$_2$CO for pointing positions that were fit after constraining the absorption velocities. }
	\label{tab:H2CO_EW_NS_after_constrain}
	\begin{tabular}{lrrrrr} 
		\hline
		Position    & S$_\nu$~~~~~~~ & RMS          & V$_{LSR}$~~~~~      & Width~~         &$\int$S$_\nu$dv~~~~~~\\
		 ($\Delta$RA, $\Delta$Dec)          & (Jy)~~~~~~  & (Jy)~        &(km s$^{-1}$)~~~ & (km s$^{-1}$) & (Jy km s$^{-1}$)\\
		\hline
NGC7538 IRS1      & 0.074(0.005)     & 0.006       & $-$57.3(0.1)      & 2.3(0.2)   & 0.18(0.03)\\
                    & $-$0.020(0.003)  & 0.006        & $-$55.91$^{\ast}$      & 15.8(1.8)  & $-$0.35(0.09)\\

(27\arcsec,0)       & 0.045(0.005)     & 0.008      & $-$57.43(0.12)      & 2.2(0.3)   & 0.10(0.03)\\
                    & $-$0.039(0.004)  & 0.008        & $-$55.59$^{\ast}$      & 7.7(0.6)   & $-$0.32(0.05)\\
($-$27\arcsec,0)    & 0.063(0.004)     & 0.006         & $-$57.75(0.07)      & 2.3(0.2)   & 0.16(0.02)\\
                    & $-$0.010(0.002)  & 0.006         & $-$56.24$^{\ast}$    & 17.7(3.3)  & $-$0.18(0.07)\\
($-$54\arcsec,0)    & 0.029(0.003)     & 0.004         & $-$58.39(0.12)      & 2.6(0.3)   & 0.08(0.02)\\
                    & $-$0.012(0.002)  & 0.004        & $-$56.56$^{\ast}$  & 10.0(1.5)   & $-$0.12(0.04)\\
(0,27\arcsec)       &0.016(0.005)       &0.007       &$-$57.49(0.30)        &2.3(0.8)   & 0.04(0.03)\\
                    & $-$0.011(0.002)  & 0.007        & $-$55.57$^{\ast}$      & 19.7(3.3)   & $-$0.24(0.08)\\
(0,$-$27\arcsec)    & 0.076(0.005)     & 0.006         & $-$57.28(0.06)      & 2.5(0.2)   & 0.20(0.03)\\
                    & $-$0.044(0.004)  & 0.006        & $-$56.25$^{\ast}$      & 8.6(0.6)   & $-$0.41(0.07)\\
(0,$-$54\arcsec)    & 0.075(0.032)     & 0.005         & $-$56.24(0.15)      & 4.5(0.7)   & 0.36(0.21)\\
                    & $-$0.078(0.033)  & 0.005        & $-$56.60$^{\ast}$      & 7.7(0.9)   & $-$0.64(0.34)\\
		\hline
        \multicolumn{6}{p{10cm}}{$^{\ast}$ Fixed parameter in the fit, thus, no uncertainty reported. All other uncertainties are 1$\sigma$ statistical errors from the fit. }\\		
\end{tabular}
\end{table*}

\begin{table*}
	\centering
	\caption{Line parameters of 12.2$\,$GHz CH$_3$OH observations.}
	\label{tab:CH30H}
	\begin{tabular}{lrrrrr} 
		\hline
		Position    & S$_\nu$~~~~~~~ & RMS          & V$_{LSR}$~~~~~      & Width~~         &$\int$S$_\nu$dv~~~~~~\\
		 ($\Delta$RA, $\Delta$Dec)          & (Jy)~~~~~~  & (Jy)~        &(km s$^{-1}$)~~~ & (km s$^{-1}$) & (Jy km s$^{-1}$)\\
		\hline
   NGC7538 IRS1     & 56.45(0.64)   & 0.01      & $-$56.071(0.003)         & 0.55(0.01)            & 32.87(0.83)\\
		            & 11.57(0.28)   & 0.01      & $-$57.17(0.04)         & 2.51(0.08)            & 30.91(1.66)\\
		            & 20.86(0.42)   & 0.01      & $-$61.07(0.01)         & 0.92(0.02)            & 20.35(0.88) \\
(108\arcsec,0)      & 0.06(0.01)   & 0.01      & $-$56.09(0.04)        & 0.58(0.12)           & 0.04(0.01)   \\
                    & $-$0.02(0.01) & 0.01     & $-$56.18(0.30)        & 4.09(0.81)           & $-$0.11(0.04)   \\ 
(54\arcsec,0)       & 6.9(0.1)      & 0.01     & $-$56.073(0.003)        & 0.55(0.01)           & 4.01(0.10)      \\
                    & 1.29(0.03)     & 0.01     & $-$57.20(0.05)         & 2.46(0.09)            & 3.38(0.21)     \\
                    & 2.52(0.05)     & 0.01     & $-$61.07(0.01)         & 0.91(0.02)            & 2.45(0.11)    \\
(27\arcsec,0)       & 35.4(0.4)      & 0.01     & $-$56.072(0.003)         & 0.55(0.01)            & 20.58(0.52)   \\
                    & 7.17(0.17)     & 0.01     & $-$57.18(0.04)         & 2.51(0.08)            & 19.12(1.03)  \\
                    & 12.99(0.26)    & 0.01     & $-$61.07(0.01)         & 0.92(0.02)            & 12.66(0.54)   \\
($-$27\arcsec,0)    & 32.20(0.40)    & 0.01     & $-$56.070(0.003)         & 0.55(0.01)            & 18.75(0.48)   \\
                    & 6.59(0.16)     & 0.01     & $-$57.16(0.04)         & 2.51(0.08)            & 17.57(0.95)  \\
                    & 11.93(0.24)    & 0.01     & $-$61.07(0.01)         & 0.92(0.02)            & 11.6(0.5)  \\
($-$54\arcsec,0)    & 4.17(0.05)     & 0.01     & $-$56.069(0.003)         & 0.55(0.01)            & 2.42(0.06)    \\
                    & 0.83(0.02)     & 0.01     & $-$57.15(0.04)         & 2.43(0.08)            & 2.14(0.12)      \\
                    & 1.56(0.03)     & 0.01     & $-$61.07(0.01)         & 0.91(0.02)            & 1.51(0.07)    \\
($-$108\arcsec,0)   & 0.07(0.01)     & 0.01     & $-$56.14(0.03)         & 0.53(0.08)            & 0.04(0.01)    \\
(0,108\arcsec)      & 0.07(0.01)     & 0.01     & $-$56.10(0.03)         & 0.41(0.07)            & 0.03(0.01)    \\
(0,54\arcsec)       & 6.26(0.07)     & 0.01     & $-$56.062(0.003)         & 0.55(0.01)            & 3.64(0.10)    \\
                    & 1.30(0.03)     & 0.01     & $-$57.14(0.04)         & 2.48(0.07)            & 3.42(0.18)   \\
                    & 2.27(0.05)     & 0.01     & $-$61.06(0.01)         & 0.92(0.02)            & 2.21(0.10)   \\
(0,27\arcsec)       & 34.4(0.4)      & 0.01     & $-$56.067(0.003)         & 0.55(0.01)            & 20.05(0.51)  \\
                    & 7.05(0.17)     & 0.01     & $-$57.16(0.04)         & 2.5(0.1)            & 18.8(1.0)  \\
                    & 12.59(0.25)    & 0.01     & $-$61.06(0.01)         & 0.92(0.02)            & 12.27(0.53)    \\ 
(0,$-$27\arcsec)    & 31.87(0.37)    & 0.01     & $-$56.074(0.003)         & 0.55(0.01)            & 18.54(0.48)   \\
                    & 6.44(0.16)     & 0.01     & $-$57.18(0.04)         & 2.48(0.08)            & 17.02(0.95)   \\
                    & 11.85(0.24)    & 0.01     & $-$61.07(0.01)         & 0.92(0.02)            & 11.6(0.5)   \\
(0,$-$54\arcsec)    & 4.38(0.07)     & 0.01     & $-$56.080(0.003)         & 0.55(0.01)            & 2.54(0.08)    \\
                    & 0.67(0.03)     & 0.01     & $-$57.23(0.07)         & 2.06(0.14)            & 1.47(0.15)   \\
                    & 1.63(0.04)     & 0.01     & $-$61.08(0.01)         & 0.90(0.02)            & 1.57(0.08)   \\
(0,$-$108\arcsec)   & $-$0.24(0.01)  & 0.01     & $-$55.97(0.04)         & 4.8(0.1)            & $-$1.25(0.05)   \\
                    & 0.08(0.01)     & 0.01     & $-$56.06(0.04)         & 0.6(0.1)           & 0.05(0.02)   \\
                    & 0.03(0.01)     & 0.01     & $-$61.2(0.1)           & 0.68(0.23)            & 0.024(0.015)     \\
\hline
\multicolumn{6}{p{10cm}}{Line parameters obtained from Gaussian fits, 1$\sigma$ statistical errors from the fit are reported in parenthesis. The spectra were smoothed to a channel width of 0.225$\,$km s$^{-1}$.}\\	
\end{tabular}
\end{table*}

\begin{table*}
	\centering
	\caption{Line parameters of 13.044$\,$GHz SO observations.}
	\label{tab:SO}
	\begin{tabular}{lrrrrr} 
		\hline
		Position    & S$_\nu$~~~~~~~ & RMS          & V$_{LSR}$~~~~~      & Width~~         &$\int$S$_\nu$dv~~~~~~\\
		 ($\Delta$RA, $\Delta$Dec)          & (Jy)~~~~~~  & (Jy)~        &(km s$^{-1}$)~~~ & (km s$^{-1}$) & (Jy km s$^{-1}$)\\
		\hline
	NGC7538 IRS1    & 0.052(0.004) & 0.007          & $-$56.99(0.18)         & 2.8(0.3)            & 0.16(0.03)\\
(108\arcsec,0)      & \ldots       & 0.008           & \ldots                   & \ldots                  & \ldots \\
(54\arcsec,0)       & 0.052(0.003) & 0.005          & $-$56.86(0.07)        & 2.6(0.7)            & 0.15(0.02)\\
(27\arcsec,0)       & 0.055(0.005) & 0.007          & $-$57.09(0.11)         & 2.7(0.3)            & 0.16(0.03)\\
($-$27\arcsec,0)    & 0.028(0.004) & 0.007          & $-$57.22(0.23)         & 3.04(0.54)            & 0.09(0.03)\\
($-$54\arcsec,0)    & 0.008(0.002) & 0.005          & $-$57.5(0.7)         & 6.49(1.66)            & 0.06(0.03)\\
($-$108\arcsec,0)   & 0.011(0.005) & 0.006          & $-$56.58(0.41)         & 1.84(0.96)            & 0.02(0.02)\\
(0,108\arcsec)      & \ldots        & 0.008         & \ldots                   & \ldots                  & \ldots \\
(0,54\arcsec)       & \ldots        & 0.005        & \ldots                   & \ldots                  & \ldots \\
(0,27\arcsec)       & 0.018(0.004)  & 0.005        & $-$56.62(0.29)         & 2.49(0.69)            & 0.05(0.03)\\
(0,$-$27\arcsec)    & 0.067(0.004)  & 0.008        & $-$57.25(0.10)         & 3.39(0.23)            & 0.24(0.03)\\
(0,$-$54\arcsec)    & 0.075(0.002)  & 0.004        & $-$56.5(0.1)         & 4.45(0.16)            & 0.36(0.02)\\
(0,$-$108\arcsec)   & 0.057(0.003)  & 0.007        & $-$56.02(0.11)         & 3.72(0.25)            & 0.23(0.03)\\
		\hline
\multicolumn{6}{p{10cm}}{Line parameters obtained from Gaussian fits, 1$\sigma$ statistical errors from the fit are reported in parenthesis. The spectra were smoothed to a channel width of 0.491$\,$km s$^{-1}$.}\\	

	\end{tabular}
\end{table*}

\begin{table*}
	\centering
	\caption{Line parameters of the other transitions.}
	\label{tab:H2CO_CH3OH_NH3}
	\begin{tabular}{lrrrrr} 
		\hline
		Spectral Line   & S$_\nu$~~~~~~   &RMS    & V$_{LSR}$~~~~ & Width~~~ & $\int$S$_\nu$dv~~~~~\\
		                & (Jy)~~~~~       &(Jy)   & (km s$^{-1}$)~~~& (km s$^{-1}$)~  & (Jy km s$^{-1}$)\\
		\hline
1$\,$cm H$_{2}$CO       & 0.099(0.003)    & 0.007 & $-$56.68(0.04)  & 2.6(0.1)     & 0.27(0.02)\\
                        & $-$0.034(0.004) & 0.007 & $-$64.3(0.1)    & 2.04(0.25)   & $-$0.08(0.02)\\
6$\,$cm H$_{2}$CO       & 0.443(0.006)    & 0.014 & $-$60.12(0.01)  & 0.84(0.02)   & 0.40(0.01)\\
                        & 1.310(0.006)    & 0.014 & $-$57.905(0.002)& 0.94(0.01)   & 1.32(0.01)\\
                        & $-$0.218(0.002) & 0.014 & $-$55.95(0.05)  & 6.65(0.09)   & $-$1.55(0.04)\\
26.847$\,$GHz CH$_{3}$OH& $-$0.156(0.002) & 0.009 & $-$59.58(0.04)  & 5.8(0.1)     & $-$0.97(0.03)\\
28.316$\,$GHz CH$_{3}$OH& $-$0.101(0.003) & 0.008 & $-$59.7(0.1)    & 4.1(0.1)     & $-$0.44(0.03)\\
28.970$\,$GHz CH$_{3}$OH& 0.128(0.006)    & 0.007 & $-$59.28(0.05)  & 1.8(0.1)     & 0.25(0.03)\\
                        & 0.264(0.005)    & 0.007 & $-$56.26(0.02)  & 2.17(0.06)   & 0.61(0.03)\\
26.518$\,$GHz NH$_{3}$  & $-$0.179(0.003) & 0.009 & $-$59.49(0.07)  & 6.91(0.16)   & $-$1.32(0.06)\\
		\hline
\multicolumn{6}{p{12cm}}{Line parameters obtained from Gaussian fits, 1$\sigma$ statistical errors from the fit are reported in parenthesis. All spectra were smoothed to a channel width of 0.2\kms, with exception of the 6$\,$cm H$_2$CO line (0.047\kms~channel width) to fit the narrow masers.}\\	
	\end{tabular}
\end{table*}

\section{Analysis and Discussion}

\begin{figure*}
\includegraphics[trim=5cm 8cm 1.2cm 3.5cm, clip,scale =1]{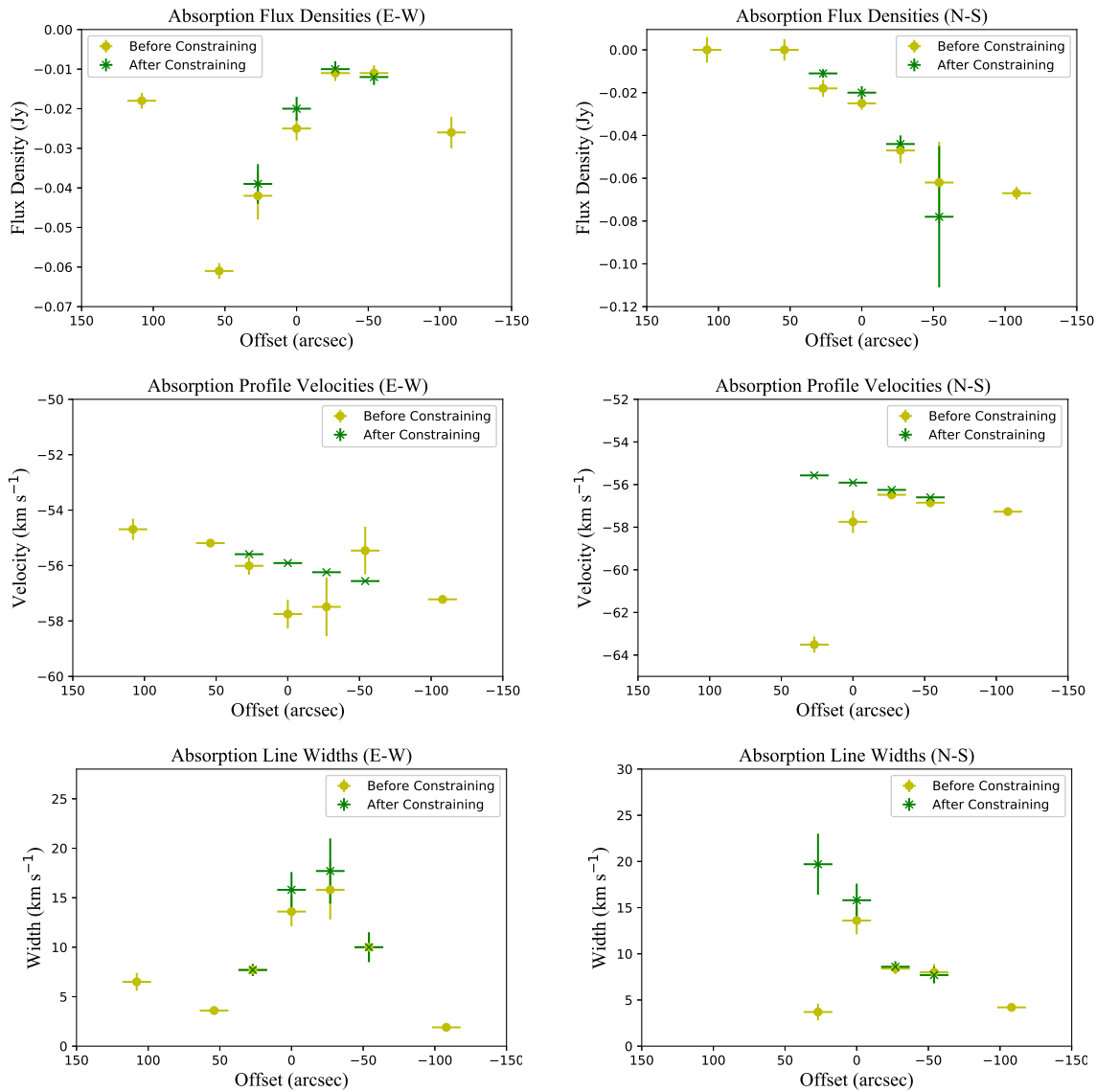}
\caption{Line parameters of 2$\,$cm H$_2$CO absorption (flux density, velocity and linewidth) from the cross-scan observations of NGC$\,$7538$\,$IRS1. Left and right panels show the East-West and North-South scans, respectively. Two different symbols are used to show the line parameters from the unconstrained (Table~\ref{tab:H2CO}) and constrained Gaussian fits (Table~\ref{tab:H2CO_EW_NS_after_constrain}). }
\label{fig:figure_H2CO_2cm_abs_EW_NS}
\end{figure*}

\begin{figure*}
\begin{tabular}{cc}
\includegraphics[trim=5cm 8cm 1.2cm 3.5cm, clip,scale =1]{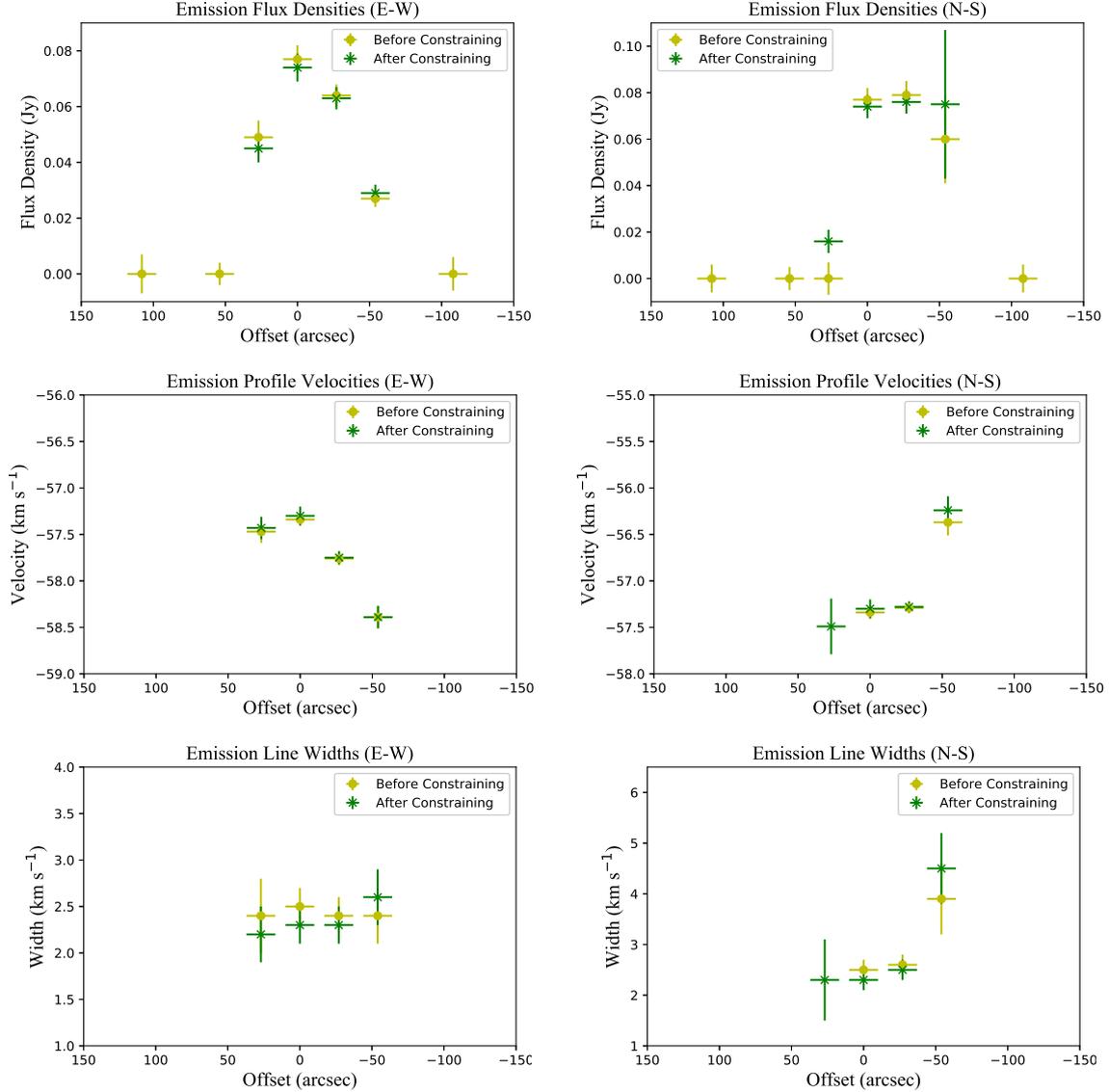}
\end{tabular}
\caption{As in Figure~\ref{fig:figure_H2CO_2cm_abs_EW_NS} but for the emission line. Emission was detected in only the central 4 pointing positions in the East-West scan, and in 4 positions in the North-South scan (see also Figure~\ref{fig:map_figure}).}
\label{fig:figure_H2CO_2cm_Ems_EW_NS}
\end{figure*}

\begin{figure*}
\includegraphics[trim=5cm 14cm 1.2cm 9cm, clip,scale =1]{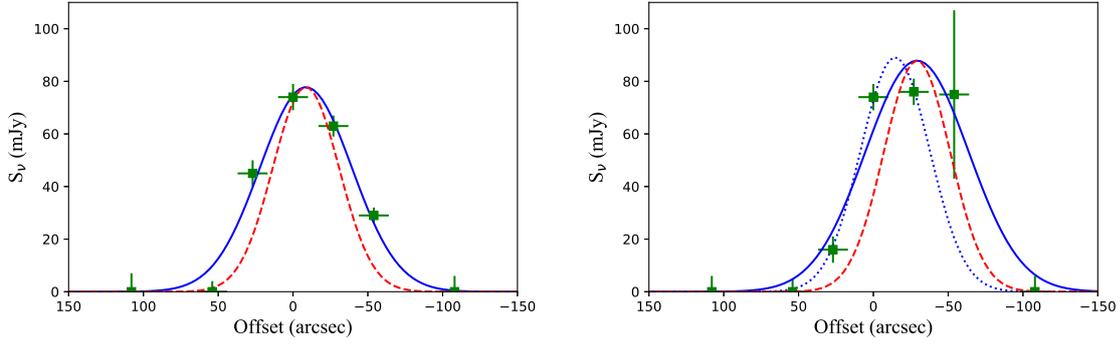}
\caption{Peak flux density distribution (East-West, left; North-South, right) of 2$\,$cm H$_2$CO emission (Table~\ref{tab:H2CO_EW_NS_after_constrain}). Continuous (blue) lines show Gaussian fits of the spatial distribution. In the right panel, we also show the Gaussian fit including relative weights due to different error bars (dotted blue line). Dashed lines in both panels show the beam pattern of the telescope centered at the peak position.}
\label{fig:figure_GaussFit_H2CO_2cm_Ems_EW_NS}
\end{figure*}

\begin{figure*}
\includegraphics[trim=5cm 14cm 1.2cm 9cm, clip,scale =1]{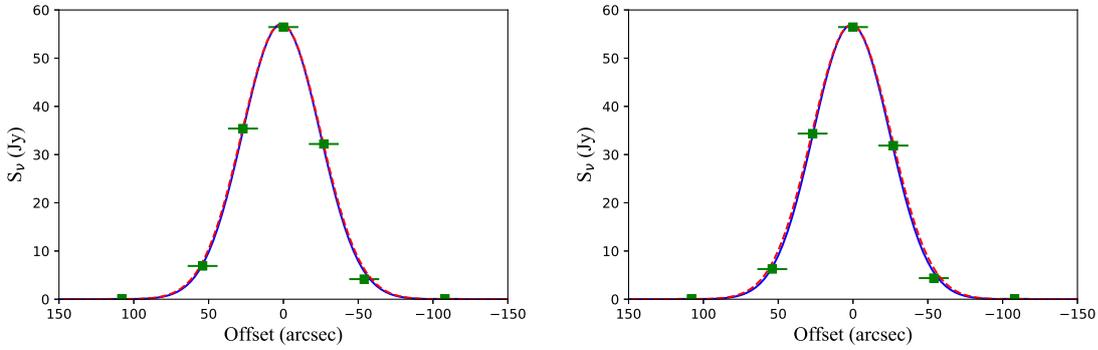}
\caption{Same as Figure~\ref{fig:figure_GaussFit_H2CO_2cm_Ems_EW_NS} but for the main peak of the 12.2$\,$GHz CH$_3$OH emission (Table~\ref{tab:CH30H}). The FWHM of the spatial distribution matches the HPBW of the telescope, i.e., the maser is unresolved. }
\label{fig:figure_GaussFit_CH3OH_12GHz_Ems_EW_NS}
\end{figure*}

\begin{figure*}
\includegraphics[trim=5cm 14cm 1.2cm 9cm, clip,scale =1]{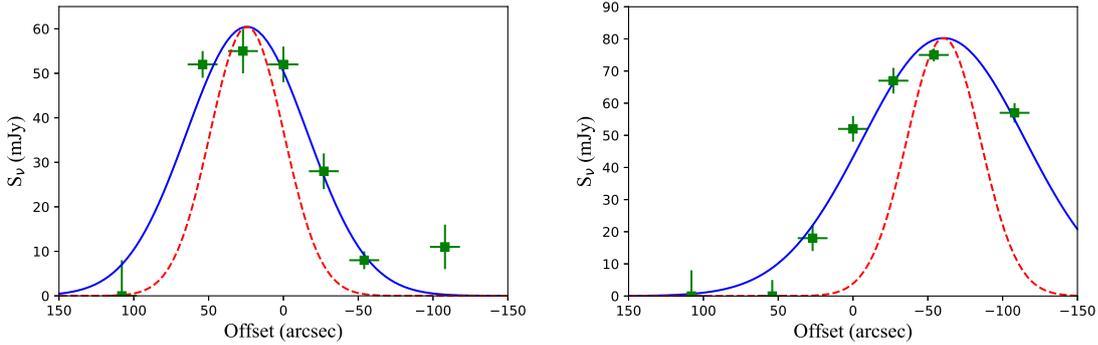}
\caption{Same as Figure~\ref{fig:figure_GaussFit_H2CO_2cm_Ems_EW_NS} but for 13.044$\,$GHz SO (Table~\ref{tab:SO}). Note that the SO emission is clearly extended.}
\label{fig:figure_GaussFit_SO_13GHz_Ems_EW_NS}
\end{figure*}

\begin{center}
\begin{figure*}
\centering
\includegraphics[width=\textwidth]{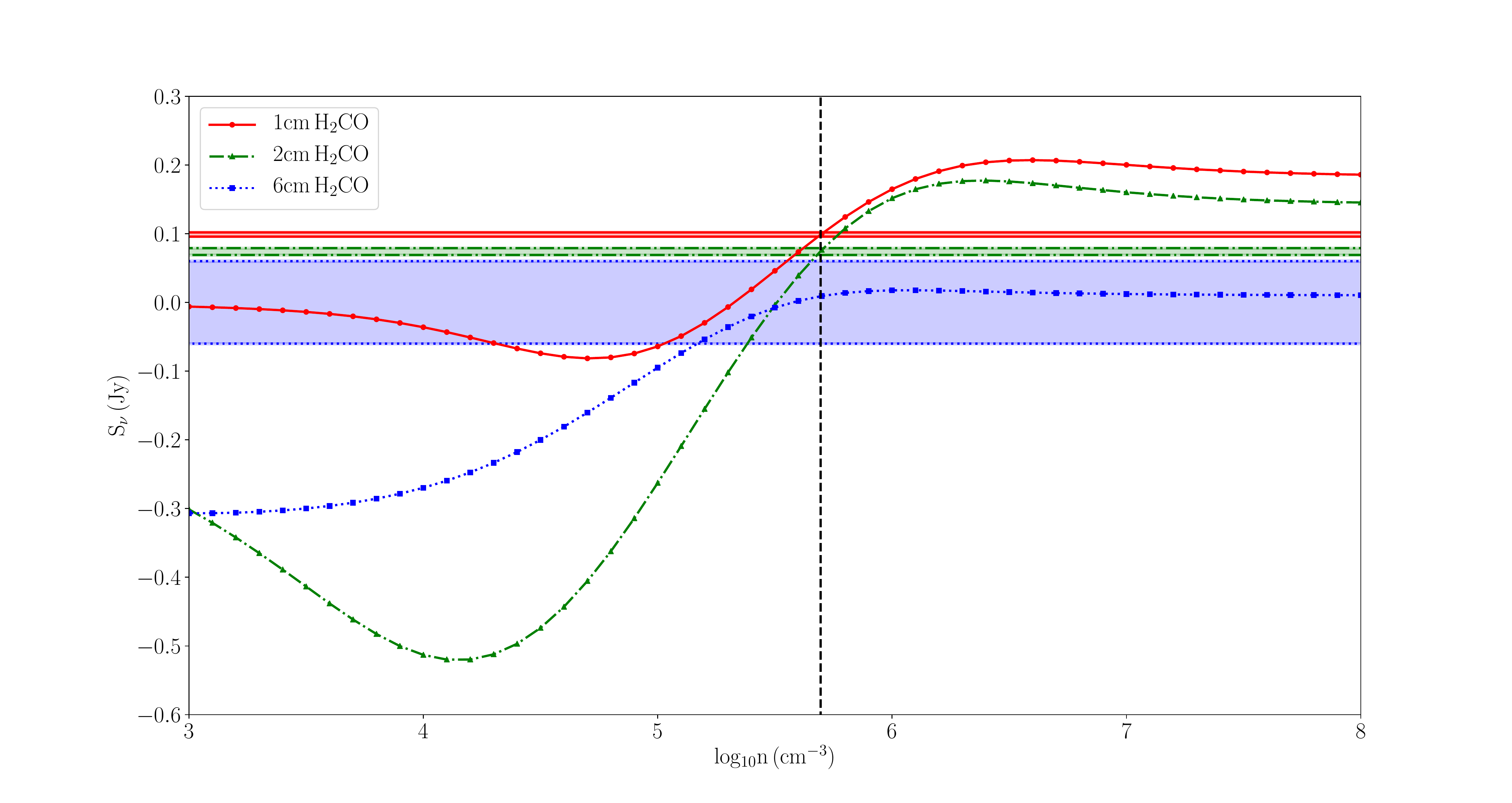}\\
\caption{Predicted flux density of \textcolor{red}{1$\,$cm (red, solid line)}, \textcolor{green}{2$\,$cm (green, dot-dashed line)} and \textcolor{blue}{6$\,$cm (blue, dotted line)} H$_2$CO lines in NGC$\,$7538$\,$IRS1 as a function of density computed with RADEX. Shaded colour regions within horizontal lines (same line style mentioned above) show the range of observed flux density values including uncertainties (see Figure~\ref{fig:figure_1cm_2cm_6cm} and discussion in Section~\ref{sec:nature_2cm_Emission}). We found that a density above $10^5\,$cm$^{-3}$ (vertical dashed line) would result in flux density values for the three transitions that are consistent with our observations.}
\label{fig:figure_H2CO_Line_parameters}
\end{figure*}
\end{center}

 Figure~\ref{fig:map_figure} shows extended 2$\,$cm H$_2$CO absorption toward the East, West, and South of NGC$\,$7538$\,$IRS1. The figure also shows 2$\,$cm emission overlapping absorption toward the center of the cross-scan. While it is clear that the absorption is extended, the extent of the emission is not as clear because of the superposition with absorption. In this section we discuss deconvolution of the emission and absorption features to measure the angular size of the 2$\,$cm H$_2$CO emission, and thus, to disentangle whether the emission is caused by a maser or a (quasi-)thermal mechanism.

\subsection{2$\,$cm H$_2$CO Absorption}\label{sec:2cm_abs}

As shown in Figure~\ref{fig:map_figure}, absorption was detected in 11 pointing positions; the two exceptions are the far North positions at (0, 108\arcsec) and (0, 54\arcsec), where neither emission nor absorption were detected. The absorption is also less prominent toward the West of NGC$\,$7538$\,$IRS1. As also shown in Figure~\ref{fig:map_figure}, the peak of the cm radio continuum from the extended ionized gas (VLA contours) is located to the North-West of NGC$\,$7538$\,$IRS1 (see also, e.g., \citealt{Ojha_2004ApJ...616.1042O}). A larger scale GBT continuum image of the region is reported in \cite{Luisi_2016ApJ...824..125L}, which shows that the diffused radio continuum emission from the H{\small II} region extends beyond a 300\arcsec~area, i.e., greater than the field of view of Figure~\ref{fig:map_figure}.

Based on the line profiles from the unconstrained-fits shown in Figure~\ref{fig:figure_H2CO_14GHz}, the pointing positions (108\arcsec,0), (54\arcsec,0), ($-$108\arcsec,0) and (0,$-$108\arcsec) are the least affected by overlapping absorption and emission lines (see Figure~\ref{fig:map_figure}). As mentioned above, in order to characterize the emission, we need to reliably deconvolve the absorption from the emission in the central pointing positions. To accomplish this, we have to characterize the absorption profiles toward the pointing positions with significant emission overlap. Comparing the velocities of the H$_2$CO absorption not affected by emission in the East-West direction, i.e., $V_{LSR}~(108\arcsec,0)$ = $-$54.69(0.38)\kms, $V_{LSR}~(54\arcsec,0)$ = $-$55.19(0.07)\kms, $V_{LSR}~(-108\arcsec,0)$ = $-$57.22(0.15)\kms ~(see Table~\ref{tab:H2CO}), we find a smooth velocity gradient across the East-West scan, with redshifted absorption toward the East and blueshifted absorption toward the West. We note that similar velocity gradients are seen at $\sim 1$\arcmin~scales, e.g., HCO$^+$ (Figure~7 of \cite{Sun_2009MNRAS.392..170S}, with the caveat that the HCO$^+$ (1-0) map is centered 1.3\arcmin~South of IRS1; see also \citealt{Sandell_2012}), CO (\citealt{Qui_2011ApJ...728....6Q}, \citealt{Sandell_2012}), CS \citep{Kameya_1986Ap&SS.118..449K}, although see the complex NH$_3$ velocity structure reported by \cite{Zheng_2001ApJ...550..301Z}.

As a way to better constrain the line parameters of the absorption, we spatially interpolated the peak absorption velocity in the central pointing positions, and fixed the value of the peak absorption in new Gaussian fits of the combined emission+absorption profiles (East-West spectra highlighted in black in Figure~\ref{fig:figure_H2CO_14GHz}, left column). As a clear H$_2$CO absorption profile was not detected toward the North pointing positions, we used the interpolated H$_2$CO absorption velocity toward the central pointing position (0,0) and the absorption velocity at (0,$-$108\arcsec), i.e., $-$57.27(0.09)\kms~(see Table~\ref{tab:H2CO}) to interpolate (and extrapolate in the case of the (0,27\arcsec) pointing) the H$_2$CO peak absorption velocities in the North-South direction, and used them as fixed parameters to fit the emission spectra (Figure~\ref{fig:figure_H2CO_14GHz}, right column). The Gaussian fits of the emission components obtained by constraining the peak absorption velocities are shown with orange solid lines in Figure~\ref{fig:figure_H2CO_14GHz}; the line parameters are listed in Table~\ref{tab:H2CO_EW_NS_after_constrain}. Inspection of the fits in Figure~\ref{fig:figure_H2CO_14GHz} shows that constraining peak absorption velocities results in detection of a weak emission line toward the (0,27\arcsec) position, and reasonable fits for all but the (0,$-54$\arcsec) pointing position, where the line profile is fit by two overlapping lines, an emission and absorption line of similar absolute peak intensities. Both the free and constrained Gaussian fits show that we cannot reliably measure the line parameters of the emission line toward (0,$-54$\arcsec), as demonstrated by the large peak flux density uncertainty for this pointing position (Table~\ref{tab:H2CO_EW_NS_after_constrain}). The (0,$-$54\arcsec) pointing includes NGC$\,$7538S, which shows a complex multi-peak emission profile in H$_2$CO rotational transitions (e.g., \citealt{Sandell_2010ApJ...715..919S}), which likely contributes to the observed complex line profile.

A graphical view of the absorption line parameters obtained from the unconstrained (Table~\ref{tab:H2CO}) and constrained (Table~\ref{tab:H2CO_EW_NS_after_constrain}) fits is shown in Figure~\ref{fig:figure_H2CO_2cm_abs_EW_NS}. While some differences between the two fits are evident (e.g., line velocities), both sets of absorption line parameters show a consistent description of the molecular cloud traced by 2$\,$cm H$_2$CO. For instance, the line parameters obtained from both methods show that the linewidth of the absorption in the East-West scan is greater toward the central pointing position where active star formation is taking place (e.g., \citealt{Moscadelli_Goddi_2014A&A...566A.150M}). The greater linewidth could be due to greater turbulence at the star formation site and/or to the molecular outflow in NGC$\,$7538$\,$IRS1 (e.g., \citealt{Wright_2014ApJ...796..112W}).

\subsection{Line Parameters and Angular Size of the 2$\,$cm H$_2$CO Emission Region}\label{sec:2cm_linepar_angsize}

As discussed below, our observations show that the 2$\,$cm H$_2$CO emission has a much wider spatial distribution than the 6$\,$cm H$_2$CO masers imaged by \cite{Hoffman_2003ApJ...598.1061H} and the multiple disk system around massive YSOs in the central region of IRS1 as shown by VLBI observations \citep{Moscadelli_Goddi_2014A&A...566A.150M}. Table~\ref{tab:H2CO_EW_NS_after_constrain} shows the line parameters of the emission lines after fixing the peak velocity of the absorption in each pointing position. The LSR velocity of the emission line at the center position from the constrained fit is $-$57.3(0.1)\kms, which agrees within the errors with the velocity of $-$57.34(0.07)\kms~obtained from the unconstrained fit (Table~\ref{tab:H2CO}). This peak velocity also agrees with the systemic velocity from C$^{17}$O and C$^{34}$S transitions reported by \cite{van_der_Tak_2000ApJ...537..283V}.

Figure~\ref{fig:figure_H2CO_2cm_Ems_EW_NS} shows the line parameters from the free (Table~\ref{tab:H2CO}) and constrained (Table~\ref{tab:H2CO_EW_NS_after_constrain}) fits for the emission line. As demonstrated in the figure, the line parameters of the emission line are not greatly modified by constraining the absorption velocities. This indicates that the line parameters of the emission are reliably measured, with the exception of the (0,$-54$\arcsec) pointing. We find that the linewidth is approximately uniform across the four East-West pointing positions, which indicates that the emission line is tracing a relatively uniform turbulent environment. We also found an East-West velocity gradient (Figure~\ref{fig:figure_H2CO_2cm_Ems_EW_NS}). At smaller scales than our cross-scan ($\sim 40$\arcsec), \cite{Wright_2014ApJ...796..112W} find a CO outflow centered on NGC$\,$7538$\,$IRS1 that is red-shifted toward the South-East and blue-shifted toward the North-West; a complete half-beam spacing mapping of the 2$\,$cm H$_2$CO emission in the region is necessary to further investigate the nature of the observed velocity gradient and its possible connection with the outflow in NGC$\,$7538$\,$IRS1, rotation or infall. 

As shown in the upper panels of Figure~\ref{fig:figure_H2CO_2cm_Ems_EW_NS}, the 2$\,$cm H$_2$CO emission peaks near the center pointing position, i.e., (0,0). In Figure~\ref{fig:figure_GaussFit_H2CO_2cm_Ems_EW_NS} we show the emission peak flux density values from Table~\ref{tab:H2CO_EW_NS_after_constrain} for the East-West and North-South scans (Figure~\ref{fig:map_figure}) fitted with Gaussian distributions. As the figure demonstrates, a Gaussian function reproduces reasonably well the East-West distribution of H$_2$CO emission. The FWHM of the 2$\,$cm H$_2$CO spatial distribution is 72\arcsec $\pm$ 6\arcsec~in the East-West direction. Due to the large uncertainty in the peak emission at (0,$-54$\arcsec), the angular size from the North-South scan is not well-constrained. The fit of the data points (without relative weights due to different error bars) results in a FWHM of 81\arcsec $\pm$ 12\arcsec~in the North-South direction (solid blue fit, Figure~\ref{fig:figure_GaussFit_H2CO_2cm_Ems_EW_NS}), however, when relative weights are included, the FWHM is 54\arcsec $\pm$ 6\arcsec~(blue dotted line, Figure~\ref{fig:figure_GaussFit_H2CO_2cm_Ems_EW_NS}). Assuming a Gaussian GBT beam of 52\arcsec $\pm$ 1\arcsec, the deconvolved angular size (at half maximum; assuming Gaussian distributions) of the 2$\,$cm H$_2$CO emission is 50\arcsec $\pm$ 8\arcsec~in the East-West direction. In the North-South, we obtained a FWHM of 62\arcsec $\pm$ 16\arcsec~without data weights, however, the large uncertainty in the (0,$-54$\arcsec) emission flux density precludes us from obtaining a reliable source size in the North-South direction. Nevertheless, our data reliably show that the 2$\,$cm H$_2$CO emission region is extended ($\sim 50\arcsec$) at least in the East-West direction. A complete (half-beam spacing) map of the 2$\,$cm H$_2$CO line, instead of a cross-scan, is needed to better characterize the absorption and thus, better deconvolve the emission from the absorption. As pointed out by \cite{McCauley_2011ApJ...742...58M}, mapping is necessary to obtain reliable source sizes for densitometry studies using H$_2$CO K-doublets.

Our simultaneous cross-scan observations of 12.2$\,$GHz CH$_3$OH and 13.044$\,$GHz SO allow us to conduct an independent test of the reliability of the H$_2$CO angular size determination. Figures~\ref{fig:figure_GaussFit_CH3OH_12GHz_Ems_EW_NS} and \ref{fig:figure_GaussFit_SO_13GHz_Ems_EW_NS} are as Figure~\ref{fig:figure_GaussFit_H2CO_2cm_Ems_EW_NS} but for the 12.2$\,$GHz CH$_3$OH and 13.044$\,$GHz SO cross-scans, respectively. In the case of 12.2$\,$GHz CH$_3$OH, it is well-known that the emission is caused by compact high-brightness temperature masers located within 1\arcsec~of NGC$\,$7538$\,$IRS1, e.g, \cite{Minier_2002A&A...383..614M}. We note that our 12.2$\,$GHz CH$_3$OH spectrum of NGC$\,$7538$\,$IRS1 (Figure~\ref{fig:figure_CH3OH_12GHz_EW_NS}) shows the same line profile as the VLBI spectrum shown in figure~2 of \cite{Minier_2002A&A...383..614M}, albeit with a different flux density, which is not surprising as multiple maser species are known to be variable in NGC$\,$7538$\,$IRS1, e.g., \cite{Andreev_2017ApJS..232...29A}. Given the compact distribution of 12.2$\,$GHz CH$_3$OH masers in the region, we expect our cross-scan to result in an unresolved source size. The Gaussian spatial fits of the CH$_3$OH data give a FWHM of 60\arcsec $\pm$ 1\arcsec~(average of the East-West and North-South FWHM; Figure~\ref{fig:figure_GaussFit_CH3OH_12GHz_Ems_EW_NS}). Given the GBT beam of 62\arcsec $\pm$ 1\arcsec~at 12.2$\,$GHz, this implies an unresolved source, 
i.e., the masers are unresolved, as expected. The case of 13.044$\,$GHz SO is the opposite extreme: as shown in Figure~\ref{fig:figure_GaussFit_SO_13GHz_Ems_EW_NS} the emission is extended, with a deconvolved size of 75\arcsec~(East-West) and 114\arcsec~(North-South), and the emission is not centered at NGC$\,$7538$\,$IRS1. These CH$_3$OH and SO tests validate our deconvolution strategy. 

We conclude that the 2$\,$cm H$_2$CO emission in NGC$\,$7538$\,$IRS1 is resolved; it is tracing a molecular core of $\sim$50\arcsec~FWHM size ($\sim 0.7\,$pc) in the East-West direction, and possibly narrower in the North-South direction. This physical size is more extended than the hypercompact H{\small II} region near NGC$\,$7538$\,$IRS1 \citep{Gaume_1995ApJ...438..776G} and than the molecular core assumed by \cite{McCauley_2011ApJ...742...58M}, i.e., 16\arcsec; while it is similar to the size of the millimeter and sub-millimeter dust core reported by \cite{Sandell_2004ApJ...600..269S}  (e.g., see their figure~3), and more compact than the NH$_3$ (1,1) molecular cloud imaged by \cite{Keown_2019ApJ...884....4K}. The physical size we measured is similar to the extent of the 2$\,$cm H$_2$CO emission filament-like structure in OMC-1 \citep{Bastien_1985A&A...146...86B}. In summary, our observations clearly show that the 2$\,$cm H$_2$CO emission does not originate from the sub-arcsecond 6$\,$cm maser region \citep{Hoffman_2003ApJ...598.1061H}, but from an extended molecular cloud.

\subsection{Nature of the 2$\,$cm H$_2$CO Emission}\label{sec:nature_2cm_Emission}

Based on the angular size of the 2$\,$cm H$_2$CO emission region discussed above, we obtain a peak brightness temperature of $0.33\,$K (assuming a symmetric Gaussian distribution of FWHM equal to the East-West major axis), which is consistent with optically thin thermal emission. 

The non-detection of 2$\,$cm H$_2$CO emission by \cite{Hoffman_2003ApJ...598.1061H} is consistent with the angular size and brightness temperature values reported here. \cite{Hoffman_2003ApJ...598.1061H} employed the VLA CnB configuration for which the shortest baselines were 65$\,$m, corresponding to a largest angular scale of $\approx 60$\arcsec at 2$\,$cm. The 2$\,$cm emission that we describe here would have been significantly filtered out by the interferometer. Moreover, the RMS brightness temperature of the 2$\,$cm H$_2$CO VLA observations reported by \cite{Hoffman_2003ApJ...598.1061H} is $\sim 10\,$K, which is much greater than the peak brightness temperature we obtained from the GBT observations (0.33$\,$K), thus, the extended region would have been resolved out.

As shown in Figure~\ref{fig:figure_1cm_2cm_6cm} (see also Table~\ref{tab:H2CO_CH3OH_NH3}), we detected 1$\,$cm H$_2$CO emission in NGC$\,$7538$\,$IRS1. Given the similar line profiles of the 2$\,$cm and 1$\,$cm lines, it is likely that the 1$\,$cm transition is also due to thermal emission (see profiles of the 6$\,$cm H$_2$CO masers with respect to the 1$\,$cm and 2$\,$cm lines in Figure~\ref{fig:figure_1cm_2cm_6cm}). Assuming LTE excitation of the 1 and 2$\,$cm lines and optically thin emission, we can estimate the excitation temperature of the gas by using the ratio of population levels (e.g., see equations A5 and A11 in \citealt{Araya_2005ApJS..157..279A}). We estimate an LTE excitation temperature of $\sim 20\,$K, which is similar to the value obtained for OMC-1 \citep{Bastien_1985A&A...146...86B}.  

To further explore the physical conditions that can lead to thermal emission of the 2$\,$cm and 1$\,$cm lines, without detectable emission of a thermal line in the 6$\,$cm transition, we modeled our multi-transition data using RADEX, which is a one dimensional non-LTE radiative transfer code developed by \cite{van_der_Tak_2007A&A...468..627V}. We explored a range of densities between $n_{H_2} = 10^3$ to $10^8\,$cm$^{-3}$ (with steps of 0.1 in $\log_{10}$($n(H_2)$), ortho-H$_2$CO column densities ($N_{o-H_2CO}$) between $10^{13}$ to $10^{15}\,$cm$^{-2}$ (with 10 steps per decade, and finer steps when converging to the final solution), kinetic temperatures between 10 to 500$\,$K (in steps of $10\,$K), and assumed a simple characterization of the background radio continuum consisting of Cosmic Microwave Background and free-free emission from \cite{Luisi_2016ApJ...824..125L} with a turnover frequency near 8 GHz (e.g., \citealt{Akabane_1991PASAu...9..118A}). Refining the model would require mapping the molecular lines and the background radio continuum at the frequencies of interest to determine the appropriate beam-filling factors.

Figure~\ref{fig:figure_H2CO_Line_parameters} shows the results of the model for the three K-doublet transitions included in this work as a function of molecular density at $T_{K} =$ 40$\,$K and $N_{o-H_2CO} = 1.5\times10^{14}\,$cm$^{-2}$. We show with a solid line (red), dot-dashed line (green), and dotted line (blue) the predicted flux densities for the 1$\,$cm, 2$\,$cm, and 6$\,$cm H$_2$CO lines, respectively. With parallel horizontal lines (same line style patterns and colours as above), we show the range of peak flux densities based on our observations. For the 2$\,$cm and 1$\,$cm H$_2$CO transitions, the range is given by the uncertainties listed in Tables~\ref{tab:H2CO_EW_NS_after_constrain} and \ref{tab:H2CO_CH3OH_NH3}. In the case of the 6$\,$cm H$_2$CO line (Figure~\ref{fig:figure_1cm_2cm_6cm}), the absorption must be caused by the extended molecular cloud also traced by 2$\,$cm absorption (Figure~\ref{fig:map_figure}), while the emission lines are caused by very compact masers, e.g., \cite{Hoffman_2003ApJ...598.1061H}. We inspected the 6$\,$cm H$_2$CO line profile, and concluded that a line asymmetry due to the material traced by 2$\,$cm and 1$\,$cm H$_2$CO emission is undetectable at a level of $\pm 60\,$mJy, which is the range shown with horizontal dotted lines (blue) in Figure~\ref{fig:figure_H2CO_Line_parameters}. We find that a gas density of $\sim 10^{5.7}\,$cm$^{-3}$ (vertical dashed line in Figure~\ref{fig:figure_H2CO_Line_parameters}) is consistent with our detection of 2$\,$cm and 1$\,$cm H$_2$CO emission and non-detection of 6$\,$cm thermal emission. This density is similar to the density obtained from the 1$\,$cm and 48$\,$GHz H$_2$CO lines ($n_{H_2} = 10^{5.78}\,$cm$^{-3}$) by \cite{McCauley_2011ApJ...742...58M}. We note that \cite{McCauley_2011ApJ...742...58M} assumed a source size of 16\arcsec, while our cross scan shows that the region traced by 2$\,$cm H$_2$CO emission is significantly more extended (Section~\ref{sec:2cm_linepar_angsize}).

The ortho-H$_2$CO column density of our model ($1.5\times10^{14}\,$cm$^{-2}$) is similar to the value reported by \cite{McCauley_2011ApJ...742...58M} ($N_{o-H2CO}/\Delta v = 10^{13.75}\,$cm$^{-2}$(\kms)$^{-1}$, see their Table~4), although the temperature in our model is lower than what they assumed (40$\,$K vs 220$\,$K). We find this temperature difference reasonable because the 2$\,$cm H$_2$CO emission likely traces a more extended region than that seen at the higher energy 48$\,$GHz H$_2$CO transition.\footnote{ \cite{McCauley_2011ApJ...742...58M} caution that their assumed temperature may not be correct but that the density determination is not greatly affected by temperature.} We note that \cite{Mitchell_1990ApJ...363..554M} reported evidence for a two-temperature environment in NGC$\,$7538$\,$IRS1 (25$\,$K and 176$\,$K components). \cite{McCauley_2011ApJ...742...58M} argued that the 1$\,$cm H$_2$CO absorption detected in their spectrum (see also our absorption detection in Figure~\ref{fig:figure_1cm_2cm_6cm}) and the 2$\,$cm H$_2$CO absorption reported by \cite{Hoffman_2003ApJ...598.1061H} originate from the low temperature component. Our analysis is consistent with detection of high-density and warm-temperature molecular material, i.e., the transition region between the cold extended component (traced by 6$\,$cm and 2$\,$cm H$_2$CO absorption) and the NGC$\,$7538$\,$IRS1 hot and dense core traced by higher excitation transitions.

Although we use a simple non-LTE model, we can reliably conclude that the 2$\,$cm H$_2$CO line in NGC$\,$7538$\,$IRS1 traces quasi-thermal emission, not a variable maser. We note that more complex models can be proposed to explain the results reported in this article. For example, several 2$\,$cm masers separated by a few arcseconds would look like an extended source in our GBT cross-scan. However, such masers would have to show simultaneous variability (to explain the non-detection by \citealt{Hoffman_2003ApJ...598.1061H}), the masers would need to have a very similar LSR velocity (otherwise the emission line would be asymmetric or multi-peaked), and a similar maser distribution would be needed for the 1$\,$cm transition (see Figure~\ref{fig:figure_1cm_2cm_6cm}). Therefore, the simplest explanation is that 2$\,$cm H$_2$CO traces extended, quasi-thermal emission that was resolved out by the VLA observations of \cite{Hoffman_2003ApJ...598.1061H}.

\section{Summary}

We present observations conducted with the GBT of the 2$\,$cm and 1$\,$cm H$_2$CO lines toward NGC$\,$7538$\,$IRS1, which complement previous observations by our group of the 6$\,$cm line. The observations were designed to investigate the nature of the 2$\,$cm H$_2$CO emission in the region, as its velocity corresponds to the velocity of 6$\,$cm H$_2$CO masers and also the systemic velocity of the thermal molecular gas. An East-West/North-South cross scan of the region revealed that the 2$\,$cm H$_2$CO emission is tracing an extended molecular core ($\sim 50$\arcsec~in the East-West direction), which implies a low brightness temperature ($\sim 0.33\,$K). In addition, LTE and non-LTE analyses, including the 6$\,$cm and 1$\,$cm data, show that the 2$\,$cm emission is not caused by a maser mechanism, but rather is a quasi-thermal line. Our analysis indicates that the 2$\,$cm H$_2$CO emission originates from a dense ($\sim 10^5$ to $10^6\,$cm$^{-3}$) and warm ($\sim 40\,$K) molecular core, which marks the transition between the lower temperature/density extended molecular cloud (traced by 6$\,$cm H$_2$CO absorption) and the hot and dense core in NGC$\,$7538$\,$IRS1 (traced by higher excitation transitions). Although this intermediate-temperature and high-density region is expected as molecular clouds collapse to form hot molecular cores, 2$\,$cm H$_2$CO emission (like the one detected in this work) is rare. It is likely that in other regions the 2$\,$cm H$_2$CO emission is engulfed (spectrally blended) with 2$\,$cm H$_2$CO absorption from the most extended and lower-density gas, preventing detection of the emission. Given that both components (absorption and emission) are extended, detection of 2$\,$cm H$_2$CO emission is challenging with interferometers, and therefore, single-dish high-angular resolution mapping is required to investigate this high-density warm-temperature transition envelope. Our work further exemplifies the potential of low K-doublet H$_2$CO observations as density probes in high-mass star forming regions as highlighted in the literature (e.g., \citealt{McCauley_2011ApJ...742...58M}).

\section*{Acknowledgements}

E.D.A. acknowledges participation of Liang Yuan (WIU alumnus) in the initial stages of the data reduction and analysis of this project, that resulted in an AAS poster presentation \citep{Yuan_2011AAS...21812904Y}. E.D.A. also acknowledges partial support from NSF grant AST-1814063, and computational resources donated by the WIU Distinguished Alumnus Frank Rodeffer. P.H. acknowledges partial support from NSF grant AST-1814011. This article is based upon work supported by the Green Bank Observatory which is a major facility funded by the National Science Foundation operated by Associated Universities, Inc. The National Radio Astronomy Observatory (NRAO) is a facility of the National Science Foundation operated under cooperative agreement by Associated Universities, Inc. We acknowledge the careful revision and suggestions of an anonymous referee, which helped us to improve this article. This research has made use of NASA's Astrophysics Data System. This research has made use of the NASA/IPAC Infrared Science Archive, which is funded by the National Aeronautics and Space Administration and operated by the California Institute of Technology.

\section*{Data Availability}

The data underlying this article are available in the article and in its online Supporting Information.





\bibliographystyle{mnras}

\bibliography{mybib}



\appendix



\bsp	
\label{lastpage}
\end{document}